\documentclass[fleqn,usenatbib]{mnras}
\usepackage{graphicx}

\usepackage[utf8]{inputenc}
\usepackage{txfonts}
\usepackage{footnote}
\usepackage{tablefootnote}
\usepackage[para,online,flushleft]{threeparttable}
\usepackage{xcolor}
\usepackage{caption,subcaption}
\definecolor{redak}{rgb}{0.9,0.15,0.05}

\usepackage{svg}
\newcommand{\orcid}[1]{\href{https://orcid.org/#1}{\includesvg[width=10pt]{orcid}}}

\defcitealias{Gilkis2025}{GL25}

\title[Blue black hole progenitors]{Hot blue progenitors of stellar-mass black holes}
\author[Gilkis \& Laplace et al.]{Avishai~Gilkis$^{\orcid{0000-0001-8949-5131}}$ $^{1}$\thanks{\href{agilkis@ast.cam.ac.uk}{agilkis@ast.cam.ac.uk}},
Eva~Laplace$^{\orcid{0000-0003-1009-5691}}$ $^{2, 3, 4}$ \thanks{\href{eva.laplace@kuleuven.be}{eva.laplace@kuleuven.be}},
Maria~R.~Drout$^{\orcid{0000-0001-7081-0082}}$ $^{5, 6}$, 
Charles~D.~Kilpatrick$^{\orcid{0000-0002-5740-7747}}$ $^{7}$, \and
Anna~J.~G.~O'Grady$^{\orcid{0000-0002-7296-6547}}$ $^{8}$
and Christopher~A.~Tout$^{\orcid{0000-0002-1556-9449}}$ $^{1}$
\\
$^{1}$ Institute of Astronomy, University of Cambridge, Madingley Road, Cambridge CB3 0HA, United Kingdom\\
$^{2}$ Institute of Astronomy, KU Leuven, Celestijnenlaan 200D, 3001 Leuven, Belgium\\
$^{3}$ Leuven Gravity Institute, KU Leuven, Celestijnenlaan 200D, box 2415 3001 Leuven, Belgium\\
$^{4}$ Anton Pannekoek Institute for Astronomy, University of Amsterdam, Science Park 904, 1098 XH Amsterdam, the Netherlands\\
$^{5}$ David A. Dunlap Department of Astronomy and Astrophysics, University of Toronto, Toronto, M5S 3H4, ON, Canada\\
$^{6}$ Observatories of the Carnegie Institution for Science, Pasadena, 91101, CA, USA\\
$^{7}$ Center for Interdisciplinary Exploration and Research in Astrophysics (CIERA), Northwestern University, Evanston, IL 60201, USA\\
$^{8}$ McWilliams Center for Cosmology and Astrophysics, Department of Physics, Carnegie Mellon University, Pittsburgh, PA 15213, USA
}

\pubyear{2026}

\begin{document}
\label{firstpage}
\pagerange{\pageref{firstpage}--\pageref{lastpage}}

\maketitle

\begin{abstract}
While the connection between massive stars and supernova explosions is well established observationally, the link between massive stars and black hole formation remains elusive. Some massive stars may collapse directly to black holes without a successful supernova, and may therefore be observed as disappearing stars. We investigate the expected photometric properties of such black hole progenitors by combining detailed single and binary stellar evolution models with physically motivated prescriptions linking pre-collapse core structure to explosion or direct collapse outcome, together with stellar atmosphere calculations, producing synthetic photometry across standard ultraviolet to near-infrared filters. Weighting by an initial mass function and empirical binary distributions, we predict both the observable distribution of progenitor brightness and colour and the rate of direct-collapse events, which we estimate to be about $0.4$ per century for a galaxy forming stars at $1\,\mathrm{M}_\odot\,\mathrm{yr}^{-1}$. We find that black hole progenitors are predominantly hot and blue at the pre-collapse stage, with many in Wolf–Rayet phases and luminous in the ultraviolet, while only a minority are red supergiants. Consequently, searches that focus primarily on red supergiants are likely to miss a substantial fraction of direct-collapse events. Monitoring campaigns that include ultraviolet-sensitive observations of nearby star-forming galaxies therefore provide a promising route to detecting disappearing massive stars, offering a direct observational probe of black hole formation. Our results provide predictions to interpret such surveys and constrain the channels that lead to black hole formation.
\end{abstract}

\begin{keywords}
binaries:~close --- stars:~black~holes --- stars:~evolution --- stars:~Wolf--Rayet --- supergiants --- supernovae:~general
\end{keywords}

\section{Introduction}
\label{sec:intro}

The existence of stellar-mass black holes (BHs) is well established \citep{Cowley1992}, and they are observed in a variety of ways. While an isolated BH has been detected through microlensing \citep{Sahu2022}, the vast majority are observed in binary systems. The first such system in which the presence of a BH was established is Cygnus X-1 \citep{Webster1972}, a high-mass X-ray binary (HMXB) system where X-ray emission originates from an accretion disc around the BH powered by gas from the companion massive star \citep{Ramachandran2025}.

For decades, HMXBs were the predominant systems in which stellar-mass BHs were identified \citep[e.g.][]{Remillard2006}, but recently, significant contributions have also been made by spectroscopic confirmation of X-ray quiet BHs in binary systems \citep[e.g.][]{Shenar2022} and by detecting gravitational waves (GWs) from merging binary BHs \citep[e.g.][]{Abbott2016}. So-called non-interacting or dormant BHs found in wide orbits through the astrometric wobble of a companion star include examples such as Gaia BH1 \citep{ElBadry2023GaiaBH1}, Gaia BH2 \citep{ElBadry2023GaiaBH2}, and Gaia BH3 \citep{Panuzzo2024GaiaBH3}, with many more predicted to be found following the next Gaia data release \citep{ElBadry2024}.

Even though they have only been available for a decade, GW detections already vastly outnumber the detections of stellar-mass BHs compared to all other methods combined, with over $200$ detections thus far \citep{LIGO2025O4}. The GW measurements provide invaluable information on the masses of merging binary BHs, presenting an opportunity to connect the properties of BHs to the structure and evolutionary history of massive stars that give birth to BHs \citep*{Laplace2025}. Yet, the exact process by which a star transitions to become a BH remains a subject of active research \citep*[for a recent review see e.g.][]{Heger2023}.

There are various theoretical predictions for the formation channels of stellar-mass BHs \citep*[e.g][]{Burrows2025}, ranging from very low-energy transients \citep*{Nadezhin1980,Lovegrove2013,Antoni2022,Antoni2023,Antoni2026} to highly energetic supernova-like events \citep*{Gilkis2014,Quataert2019}. At the more energetic end, a BH might form from the collapse of a massive star with a rapidly rotating core and give rise to an energetic supernova \citep*[e.g.][]{Gilkis2016} and possibly also a long-duration gamma-ray burst \citep[GRB;][]{Woosley1993,Hjorth2003,Stanek2003}. The coalescence of two neutron stars is another channel in which a BH may form, accompanied by a kilonova \citep[e.g.][]{Arcavi2017} or a short-duration GRB \citep{Eichler1989,Berger2014,Abbott2017}.

Rare high-energy transients such as long-duration GRBs and superluminous supernovae are associated with particular evolutionary channels, including rapidly rotating stripped progenitors and chemically homogeneous evolution in low-metallicity binaries or single stars \citep*[e.g.][]{Izzard2004,Yoon2006,Woosley2006}. Their observed volumetric rates are several orders of magnitude below the overall core-collapse supernova (CCSN) rate, implying that such channels are unlikely to account for the bulk of stellar-mass BH formation \citep[e.g.][]{Quimby2013,Prajs2017,Pessi2025}. It is therefore prudent to consider channels that occur in a non-negligible fraction of massive stars, as motivated by core-collapse physics. When a massive star reaches the end of core silicon burning, an iron-rich core forms. This cannot provide further energy by nuclear fusion and support the core against gravitational collapse. Ultimately, once it exceeds the effective Chandrasekhar mass, the iron core collapses, aided by electron captures and photodisintegration of nuclei \citep*{Woosley2002}. In less than a second, a proto-neutron star forms and whether a successful CCSN follows depends on the competition between energy deposition into the collapsing stellar material and its binding energy \citep{Janka2012}. Therefore, it might be expected that stars with more tightly bound internal structures would be more likely progenitors of BHs, and numerous hydrodynamic simulations have predicted that stars might collapse with no accompanying supernova explosion \citep{Heger2023}. A direct-collapse event, in which the star simply disappears, might then occur.

If a substantial fraction of massive stars form BHs without a bright supernova, their disappearance may be detectable through long-term monitoring of nearby galaxies. However, identifying the most promising progenitors requires guidance from stellar evolution models: which types of stars are theoretically most likely to collapse to BHs? Observational efforts have been made to find disappearing stars through optical and infrared monitoring campaigns (e.g. \citealt{Kochanek2008,Adams2017_constraints,Neustadt2021}; \citealt*{Reynolds2015}; \citealt{Jencson2022,De2026}). While these surveys are not explicitly limited to red supergiants (RSGs), in practice they are most sensitive to luminous cool stars and low-luminosity transients \citep*{Gerke2015}, potentially associated with BH formation \citep*{Lovegrove2013}, owing to their use of optical bands and expectations from single-star evolution models. Focusing on RSGs is further motivated by the apparent lack of luminous RSG progenitors of supernovae (the missing exploding RSG problem; \citealt{Smartt2009,Davies_Beasor2020}). A possible explanation is that some of these stars quietly disappear and form BHs. While RSGs are observationally attractive targets for disappearing stars, there are strong theoretical reasons to expect that they may not dominate BH formation.
\begin{enumerate}
    \item There is increasing evidence for hydrodynamical instabilities in the envelopes of RSGs before explosion, leading to high-amplitude pulsations (\citealt*{Sengupta2026}; \citealt{Laplace2026}). For high luminosity to mass ratios, these pulsations may eject the envelope shortly before explosion, resulting in a hotter effective temperature before collapse (\citealt{Yoon2010}; \citealt*{Gofman2020};  \citealt{Bronner2025,Suzuki_Shigeyama2025}), and offering an alternative explanation for the missing RSG problem.
    \item The loosely-bound envelopes of RSGs might also be ejected following mass loss by neutrino emission, creating a long-lived low-energy event \citep[e.g.][]{Lovegrove2013} and therefore a persistent infrared source that might be confused with a surviving star, while for more compact hotter stars this might be avoided as very small ejecta masses are predicted \citep{Fernandez2018}.
    \item Hot massive endpoints such as Wolf--Rayet (WR) stars have more massive cores, which might be more difficult to explode.
    \item Most massive stars live in interacting binaries \citep{Sana2012,Sana2025}, where mass transfer, common-envelope evolution, coalescence, and binary disruption can substantially modify the masses, envelope structures, rotation, and evolutionary pathways of both stars. These interactions produce a broad diversity of pre-collapse configurations and therefore a correspondingly diverse population of BH progenitors \citep{Langer2020,Laplace2021,Gilkis2025}.
\end{enumerate}
For these reasons, observational campaigns searching for BH-forming progenitors should likely target hot, blue massive stars -- though we note that theoretical predictions for the exact pre-collapse properties of BH progenitors are lacking and often consider specific progenitor channels \citep*{Schneider2021,Schneider2024,Schneider2025}, or use simplified assumptions regarding the core-collapse outcome \citep{Langer2020}.

Here we use detailed grids of single and binary stellar evolution models and a semi-analytic neutrino-driven core-collapse model to predict the photometric properties of all single and binary stars that collapse to BHs without a successful supernova. We determine where these progenitors lie in colour–magnitude space, quantify their relative contributions by spectral type, and assess the implications for disappearing-star surveys. Our results suggest that searches focused primarily on RSGs may miss the majority of BH-forming events, which we predict are predominantly from hot and blue stars.

In Section\,\ref{sec:FailedSearches} we summarise previous observational searches for failed supernovae and disappearing stars. In Section\,\ref{sec:TheyAreBlue} we describe our numerical framework, combining detailed stellar-evolution models with atmosphere and spectral libraries to predict observable properties of BH progenitors. In Section\,\ref{sec:TheFutureIsBrightBlue} we present predicted colour–magnitude distributions and event-rate estimates for disappearing stars, including their dependence on progenitor type. In Section\,\ref{sec:TheseAreTheStarsYouShouldBeLookingFor} we discuss the implications for current and future surveys and summarise our findings.

\section{Watching stars become black holes -- the story so far}
\label{sec:FailedSearches}

The formation of BHs can be probed through several observational channels. At the faint end, low-luminosity optical/infrared transients powered by partial envelope ejection during failed explosions have been predicted \citep*[e.g.][]{Lovegrove2017}. More energetic events, sometimes termed BH-supernovae, have been proposed as outcomes of weak explosions that nevertheless result in BH formation \citep[e.g.][]{EggenbergerAndersen2025}. At the most luminous extreme, long-duration GRBs and hypernovae are associated with rapidly rotating cores collapsing to BHs \citep{Woosley1993,Woosley2006} and can be detected to cosmological distances \citep{Salvaterra2015,Levan2025}. Depending on luminosity, electromagnetic (EM) signatures of BH formation may therefore be observable from the Local Group out to Gpc scales.

Non-EM messengers provide complementary constraints. Neutrino bursts directly trace core collapse and proto-compact object evolution \citep{Janka2012} and a sudden termination of neutrino emission has long been suggested as a potential indicator of BH formation (\citealt*{Burrows1988,Beacom2001}; \citealt{Walk2020}). However, current neutrino detectors are primarily sensitive to events within about $1\,\mathrm{Mpc}$ \citep{Valtonen-Mattila2023ApJ,IceCube2024ApJ}. GWs from stellar collapse are expected to be detectable only for Galactic or very nearby extragalactic events, with realistic sensitivity limits typically of order a few kpc for canonical core-collapse signals \citep{Mueller2026GWreview}. Thus, while neutrinos and GWs offer direct probes of collapse physics, their limited reach makes them rare tools for studying BH formation.

Among EM approaches, one direct method is to monitor nearby galaxies for the disappearance of massive stars. The possibility that some massive stars collapse directly to BHs without producing a luminous supernova has motivated systematic observational searches for failed supernovae \citep{Kochanek2008,Gerke2015,Adams2017_constraints,Reynolds2015,Basinger2021,Jencson2022}. These long-term monitoring programmes compare images of galaxies taken at different epochs to identify stars that vanish without a bright transient, and have  provided a few candidates. The Large Binocular Telescope (LBT) survey identified several disappearing-star candidates (\citealt{Adams2017,Neustadt2021}; \citealt*{Kochanek2024}), most notably N6946-BH1, which has been proposed as a candidate for a BH forming in the compactness peak, the progenitor mass range in which pre-collapse cores are predicted to be particularly compact and difficult to explode \citep{Sukhbold2020}. However, interpreting such events is not straightforward. Massive stars exhibit large-amplitude variability, eruptive mass loss, and dust formation episodes that can substantially reduce their optical brightness without terminal collapse. Infrared follow-up is therefore essential to test whether a star has truly vanished or is instead heavily dust-enshrouded. In the case of N6946-BH1, the debate continues over whether the observed infrared source is an accreting BH or the result of a coalescence event \citep{Beasor2024,Kochanek2024}.

One alternative explanation for such dimming events is the asymmetric coalescence of two stars, in which dense equatorial outflows can obscure the central source for years to decades \citep{KashiSoker2017}. In such merging scenarios, radiation from the central system is blocked along certain lines of sight and reprocessed by polar dust, leading to strong mid-infrared emission and suppressed optical flux. In this picture, a disappearing optical source does not necessarily imply core collapse.

Recent multi-wavelength studies have reinforced these ambiguities. \citet{Jencson2022} showed that some extreme dimming events are followed by partial re-brightening, suggesting non-terminal processes. A particularly detailed reassessment is provided by \citet{Beasor2026a}, who analysed the failed-supernova candidate M31-2014-DS1, which is consistent with a disappearing yellow supergiant \citep{De2026}, using the James Webb Space Telescope (\textit{JWST}), radio, and X-ray observations. They find persistent mid-infrared emission at the progenitor location a decade after optical fading, consistent with a heavily dust-enshrouded star. The absence of strong X-ray emission disfavours sustained accretion on to a newly formed BH, and the spectral energy distribution exhibits features indicative of asymmetric circumstellar dust. They conclude that stellar coalescence or other non-terminal eruptive events remain viable explanations. More recently, \cite{ForesToribio2026} re-analysed \textit{JWST} observations of N6946-BH1 and argued that its long-term fading and mid-infrared properties remain more consistent with a failed supernova interpretation than with a stellar merger remnant.

Similar ambiguities may apply to other proposed systems associated with failed supernovae or BH formation channels. For example, the donor star of the HMXB system NGC~300~ULX-1 was suggested to have collapsed to a BH \citep{Chene2025}, possibly signalling the formation of a BH-neutron star binary. A subsequent study by \cite{Beasor2026b} found a surviving luminous source which disfavours this interpretation.

Current disappearing-star surveys thus provide important empirical constraints, but the interpretation of individual candidates remains uncertain. Most observational efforts to date have been most sensitive to luminous cool supergiants, although multi-band searches can also probe hotter progenitors. In particular, the U-band coverage of the LBT survey provides appreciable sensitivity to the hot, blue population considered here within its nearby galaxy sample \citep{Adams2017,Neustadt2021}. Nevertheless, crowding, extinction and the absence of shorter-wavelength UV coverage can make the hottest compact stripped progenitors more challenging to identify. As a result, most candidates identified to date are cool supergiants, including RSG and yellow supergiant (YSG) candidates, although one blue supergiant (BSG) candidate has also been reported \citep{Neustadt2021}. As we show below, theoretical models predict that direct BH formation may often occur for more compact progenitors. These stars are hotter, bluer and observationally distinct from the cool supergiants that dominate current search strategies\footnote{\citet{Yoon2012} and \citet{Eldridge2013} have shown that stripped-envelope progenitors are expected to be optically faint and therefore difficult to identify in pre-explosion imaging, and the scarcity of observed Type Ib/c progenitors is broadly consistent with this expectation.}. A systematic prediction of the photometric properties of BH progenitors across evolutionary channels is therefore required to interpret disappearing-star surveys and to guide future searches.

\section{What do stars look like before imploding?}
\label{sec:TheyAreBlue}

Our predictions are based on detailed stellar-evolution models combined with atmosphere models to derive observable magnitudes and colours at the pre-collapse stage.

\subsection{Stellar-evolution endpoints}
\label{subsec:endpoints}

We base our predictions on detailed pre-collapse stellar models \citep[][hereinafter GL25]{Gilkis2025}. These models were computed with version 15140 of \textsc{mesa} \citep{Paxton2011,Paxton2013,Paxton2015,Paxton2018,Paxton2019} for non-rotating single and binary stars at three metallicities, $Z=0.014$, $0.0056$, and $0.00224$, chosen to approximately represent Milky Way (MW), Large Magellanic Cloud (LMC) and Small Magellanic Cloud (SMC) environments, respectively. The model grid spans zero-age main-sequence (ZAMS) masses $4\,{\rm M}_\odot\le M_{\rm ZAMS}\le 107\,{\rm M}_\odot$, and binary systems sample a range of initial mass ratios and orbital periods appropriate for massive stars \citep[e.g.][]{Sana2012}. Binary evolution includes Roche-lobe overflow, common-envelope evolution, coalescence products, disrupted secondaries and systems in which the surviving star continues evolving with a compact companion which can be a white dwarf, neutron star or BH. 

Stellar evolution was followed whenever feasible to iron-core collapse; otherwise models were evolved at least to the end of core carbon burning, which is sufficient to characterise pre-collapse surface properties and to classify endpoints. For models evolved to core collapse, we determine the explosion outcome using the semi-analytical neutrino-driven supernova model of \citet{Muller2016}\footnote{The mapping between pre-collapse structure and explosion outcome remains uncertain, but \citet{Maltsev2025} find that explodability predictions calibrated to the \citet{Muller2016} model agree with most of the state-of-the-art three-dimensional core-collapse simulations that they investigated, with most disagreements involving Population III stars or a rapidly rotating progenitor outside the regime considered here.}. Here we focus on the subset of endpoints that are predicted to form BHs by direct collapse, without a successful supernova.

For later interpretation we adopt the same endpoint stellar-type classification used by \citetalias{Gilkis2025}. Non-WR endpoints are classified by effective temperature and surface hydrogen fraction as RSG, YSG, BSG, and helium giant (HeG). For WR stars we follow \citetalias{Gilkis2025} and identify WR endpoints using the transformed radius criterion (\citealt*{Schmutz1989}; \citealt{Shenar2020}), with sub-types (WNh/WN/WC/WO) assigned from the surface composition. We note that not all hot stripped helium stars are expected to display classical WR-like spectra. In our terminology, the WR classification refers specifically to models satisfying the transformed-radius criterion as in \citetalias{Gilkis2025}, corresponding to stars with sufficiently dense winds to produce WR-like emission-line spectra. Hot stripped stars that do not satisfy this criterion are treated as non-WR stripped stars despite their helium-rich surfaces, and classified as BSG if they still retain some hydrogen, and as HeGs otherwise. Such objects are increasingly observed in binary systems and may represent an important population of optically faint stripped progenitors \citep{Drout2023,Gotberg2023,Ludwig2026}.

Some progenitors may undergo luminous blue variable (LBV)-like excursions shortly before collapse, potentially altering their observational appearance \citep*{Groh2013}. We do not include LBVs as a separate endpoint class in the fiducial classification, but explore the effect of possible late-time LBV-like phases in the predicted colour--magnitude distributions of BH progenitors in Appendix\,\ref{app:lbv}.

The non-inclusion of rotation in the \citetalias{Gilkis2025} models might have a few implications. At moderate rotation rates, the main relevant effect is rotationally induced mixing, which can modify the core structure and hence shift the initial-mass ranges associated with different collapse outcomes \citep{Luo2025}, similarly to variations in convective overshooting (e.g. \citealt*{Hirschi2004}; \citealt{Temaj2024,Laplace2025}). In binaries, rotation may also limit the efficiency of mass transfer as the accretor is spun up. However, disc-regulated angular-momentum transport may allow substantial mass accretion \citep{Xing2026}, for which our non-rotating models may provide a reasonable approximation. Rotation can also play an important role in the explosion mechanism itself and has been shown to alter the dynamics in 3D CCSN simulations. For the same progenitor, including rotation can change the outcome from a failed to a successful explosion, although the exact outcome is a complex function of the progenitor structure and rotation rate \citep[e.g.][]{Powell_Mueller2020}. At sufficiently rapid rotation, qualitatively different and rarer evolutionary and collapse channels might be possible, particularly at low metallicities, including rotation-powered hypernova explosions or gamma-ray bursts \citep{MacFadyen1999,Gottlieb2022}.

\subsection{Spectral libraries and synthetic photometry}
\label{subsec:synphot}

We assign a spectral energy distribution (SED) to each stellar-evolution endpoint and compute absolute AB magnitudes by convolving the SED with the relevant filter transmission curves. The implementation follows the methodology introduced by \citet{GilkisArcavi2022}, but here we (i) compute absolute magnitudes rather than applying distance corrections, (ii) adopt the AB magnitude system and (iii) do not include reddening by interstellar or circumstellar dust. Our aim is to characterise the intrinsic colour--magnitude distribution of BH-forming progenitors.

\subsubsection{Spectral libraries}

We use three classes of spectra. For cool stars we adopt the empirical Pickles library \citep{Pickles1998}, restricted to luminous (class I) spectra appropriate for evolved massive stars. For hot, non-WR stars we use atmospheres computed with \textsc{tlusty}, adopting the BSTAR2006 grid for early B stars \citep{TlustyB2007} and the OSTAR2002 grid for O stars \citep{TlustyO2003}. These grids provide spectra as a function of effective temperature and surface gravity. For WR-like spectra we use expanding-atmosphere PoWR models (\citealt*{PoWR2002}; \citealt{PoWR2003,PoWR2015}). These models are parameterised by effective temperature and transformed radius $R_\mathrm{t}$ \citep{Schmutz1989}, and we employ grids appropriate for WN \citep{Todt2015} and WC/WO stars \citep*{Sander2012}.

\subsubsection{Interpolation within libraries}

Within each spectral library we interpolate synthetic magnitudes across the grid parameters. Pickles spectra are interpolated one-dimensionally in effective temperature. \textsc{tlusty} spectra are interpolated in two dimensions over $(T_{\rm eff},\log g)$. PoWR spectra are interpolated in two dimensions over $(T_{\rm eff},R_\mathrm{t})$. Interpolation is performed on tabulated magnitudes, ensuring smooth behaviour across parameter space and avoiding discontinuities in colour--magnitude space.

\subsubsection{Smooth transitions between spectral regimes}

Rather than applying sharp temperature boundaries between spectral classes, we implement smooth transitions to avoid artificial jumps in broad-band colours. Across the transition between cool (Pickles) and B-star (\textsc{tlusty}) spectra, fluxes are linearly blended in the interval $T_{\rm eff}\approx 15$ to $16\,\mathrm{kK}$. Similarly, fluxes are blended between B-star and O-star \textsc{tlusty} grids across $T_{\rm eff}\approx 29$ to $30\,\mathrm{kK}$. Outside these narrow transition bands a single spectral regime is used exclusively. The weights vary linearly with temperature across each interval and are applied in flux space.

\subsubsection{WR assignment and WR/non-WR blending}

WR spectra are not assigned solely on the basis of temperature. Instead, we use the transformed radius to define a continuous WR weight. For $\log_{10} R_\mathrm{t} \le 1.5$ the spectrum is treated as fully WR-like, while for $\log_{10} R_\mathrm{t} \ge 1.7$ no WR contribution is included. Between these limits the WR and non-WR spectra are blended linearly in flux space. This procedure reflects the gradual emergence of WR emission features as winds strengthen and avoids sharp transitions near the WR boundary.

Within the WR regime, subtype selection follows the surface composition. Hydrogen-rich models are assigned WNL-type spectra, transitioning smoothly to hydrogen-poor WN spectra as the surface hydrogen mass fraction decreases. For hydrogen-deficient models, WC/WO spectra are adopted when carbon dominates over nitrogen at the surface. In practice, fluxes from adjacent PoWR grids are blended across the relevant composition intervals to maintain continuity in the predicted magnitudes.

\subsubsection{Absolute AB magnitudes}

For each endpoint we compute absolute AB magnitudes in the chosen filters. Library magnitudes are rescaled to the stellar radius assuming $F \propto R^2$, and blended spectra (across temperature transitions or WR mixing) are combined in flux space before converting back to magnitudes. No extinction or reddening is applied in this work.

\subsection{Population weighting}
\label{subsec:weights}

To relate individual stellar-evolution calculations to expected occurrence rates, we assign a statistical weight to each evolutionary sequence. Primary masses are drawn from a \citet{Salpeter1955} initial mass function (IMF), which means $\mathrm{d}N/\mathrm{d}M \propto M^{-2.3}$ in the massive star regime. We account for the break in slope at $0.5\,\mathrm{M}_\odot$ according to \cite{Kroupa2001} in our normalisation, so that the event rate computed in Section\,\ref{subsec:rates} is $\approx 60\%$ higher than if we had used a single-sloped IMF\footnote{We do not account for the mass dependence of the multiplicity fraction \citep{Moe2017}.}. We assume a binary fraction of $70\%$ for massive stars \citep{Sana2012, Sana2025}. For binaries we adopt a flat mass-ratio distribution and a flat-in-log period distribution. 

Each simulated combination of initial mass, mass ratio, and orbital period is assigned a weight corresponding to the integral of the assumed distributions over the parameter-space bin represented by that model. For simulations involving post-supernova evolution of secondaries, weights are propagated consistently from the initial configuration, accounting for disruption probabilities in cases involving natal kicks.

\subsection{Colour--magnitude distributions}
\label{subsec:cmds}

\begin{figure*}
    \centering
    \includegraphics[width=\linewidth]{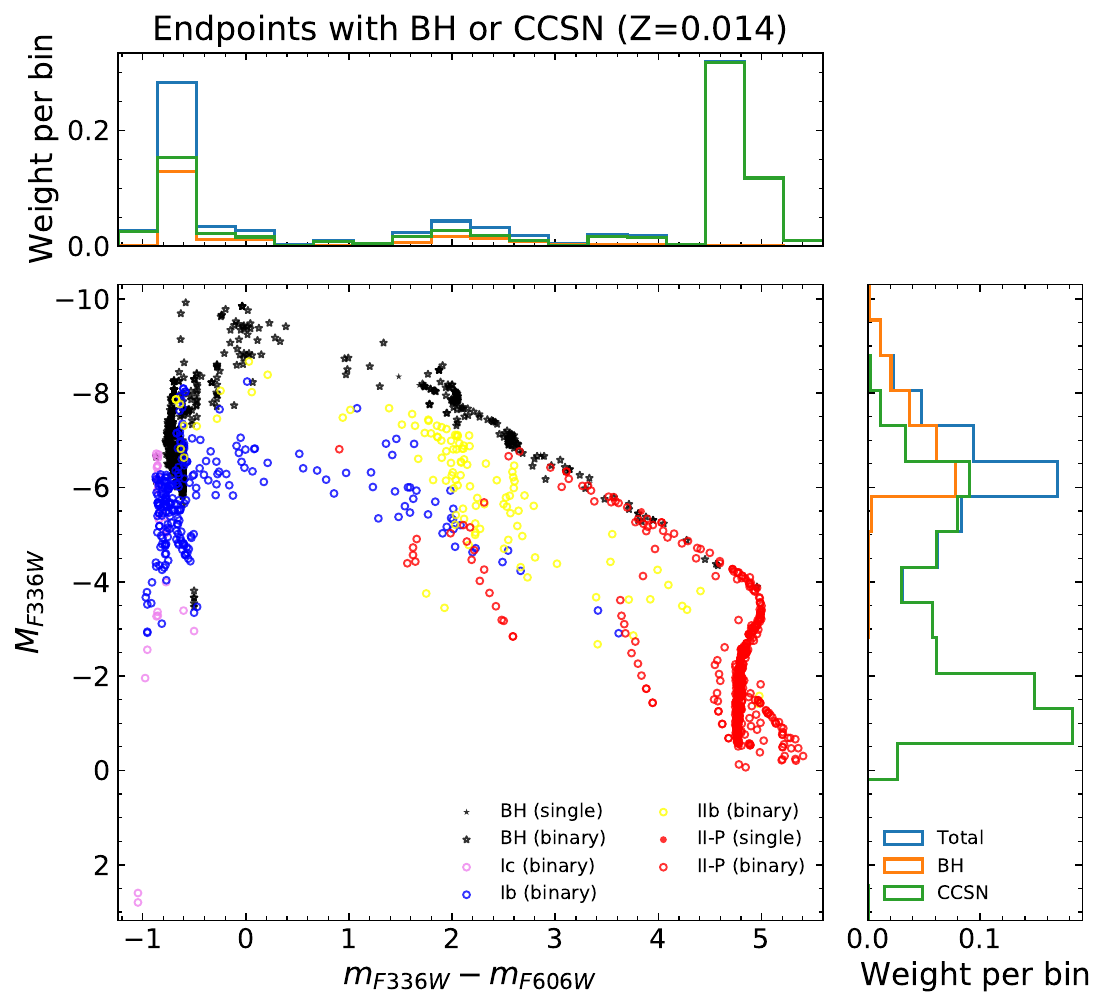}
    \caption{Colour--magnitude diagram for all endpoints predicted to produce either core-collapse supernovae or direct-collapse BHs at $Z=0.014$, including all endpoints (single stars and binary systems). We note that no dust extinction was applied.}
    \label{fig:cmd_CCSNe_plus_BHs}
\end{figure*}

\begin{figure*}
    \centering
    \begin{subfigure}[t]{0.49\textwidth}
        \centering
        \includegraphics[width=\textwidth]{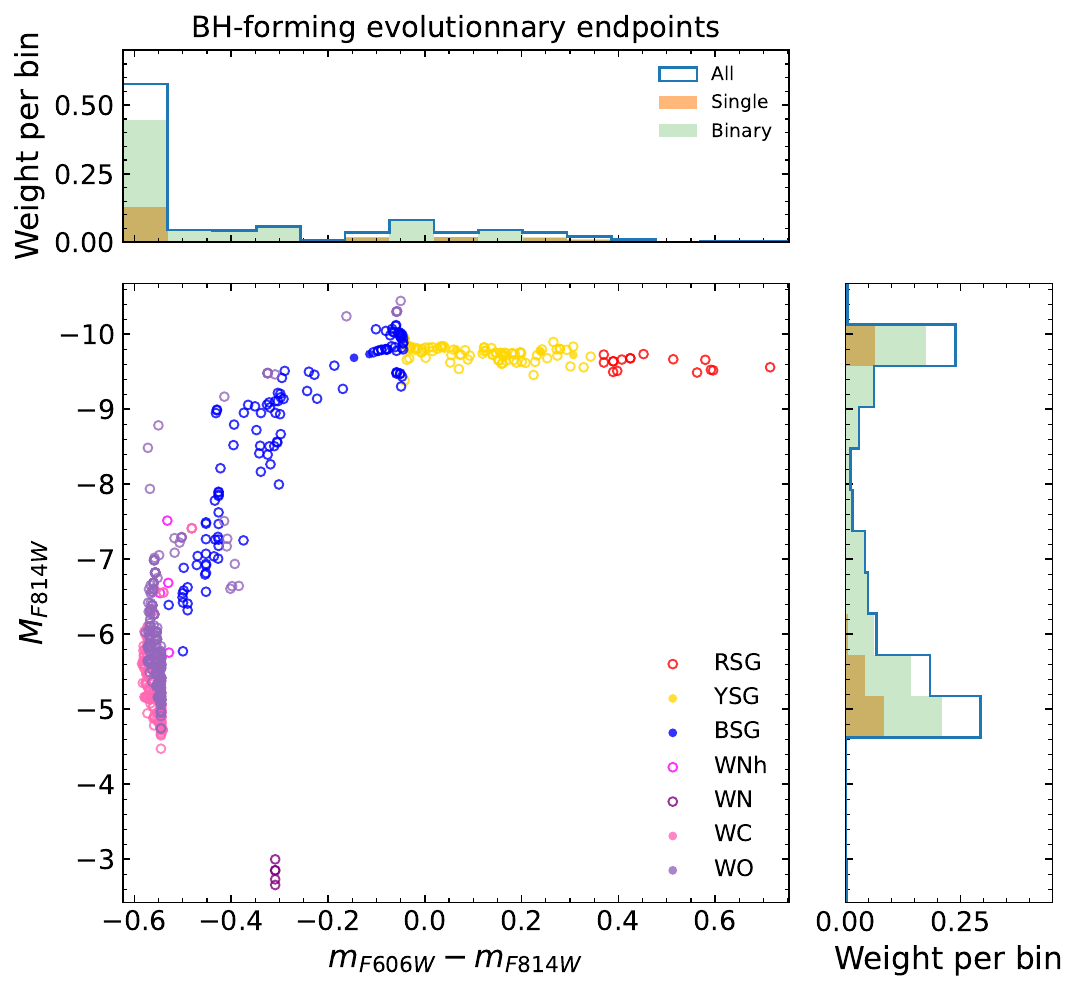}
        \caption{$m_\mathrm{F606W} - m_\mathrm{F814W}$ vs.\ $M_{\rm F814W}$}
    \end{subfigure}
    \begin{subfigure}[t]{0.49\textwidth}
        \centering
        \includegraphics[width=\textwidth]{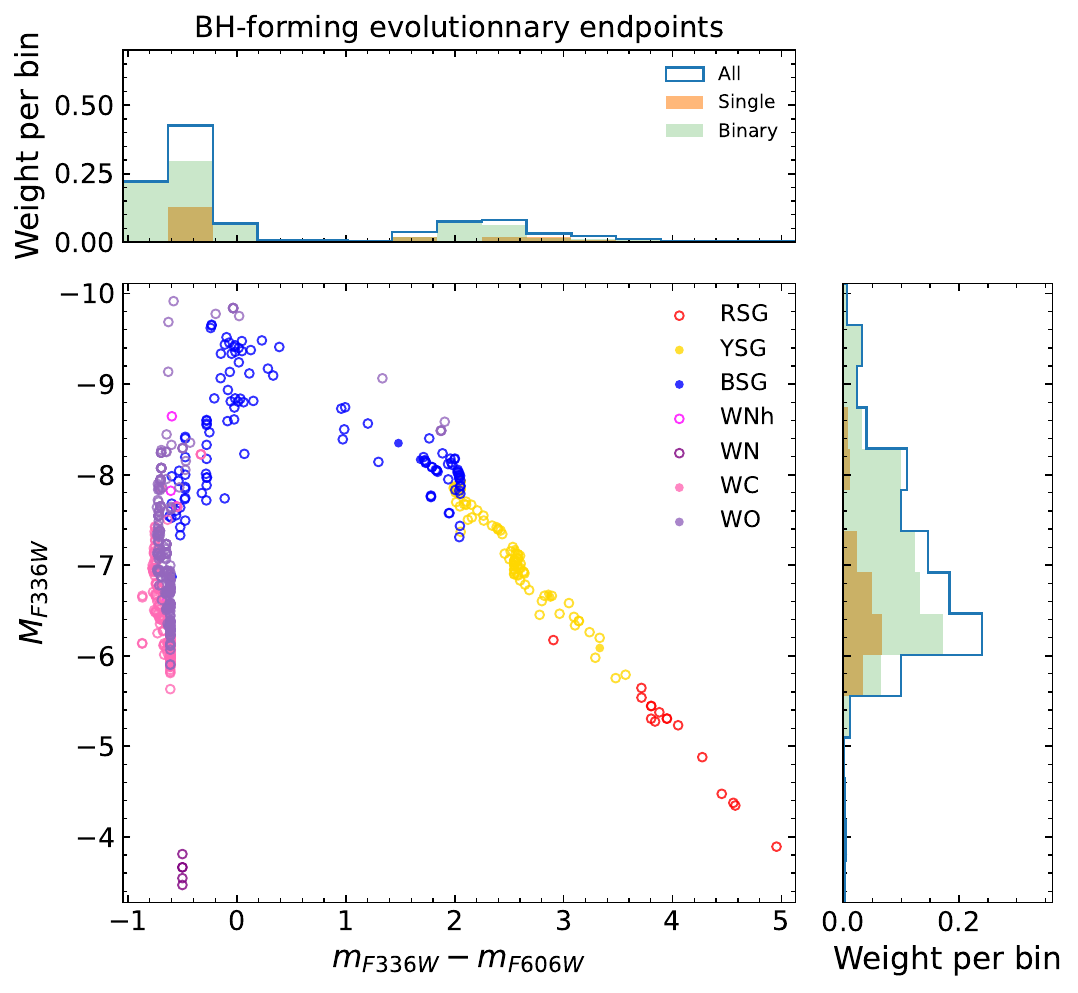}
        \caption{$m_\mathrm{F336W} - m_\mathrm{F606W}$ vs.\ $M_{\rm F336W}$}
    \end{subfigure}

    \vspace{0.8em}

    \begin{subfigure}[t]{0.49\textwidth}
        \centering
        \includegraphics[width=\textwidth]{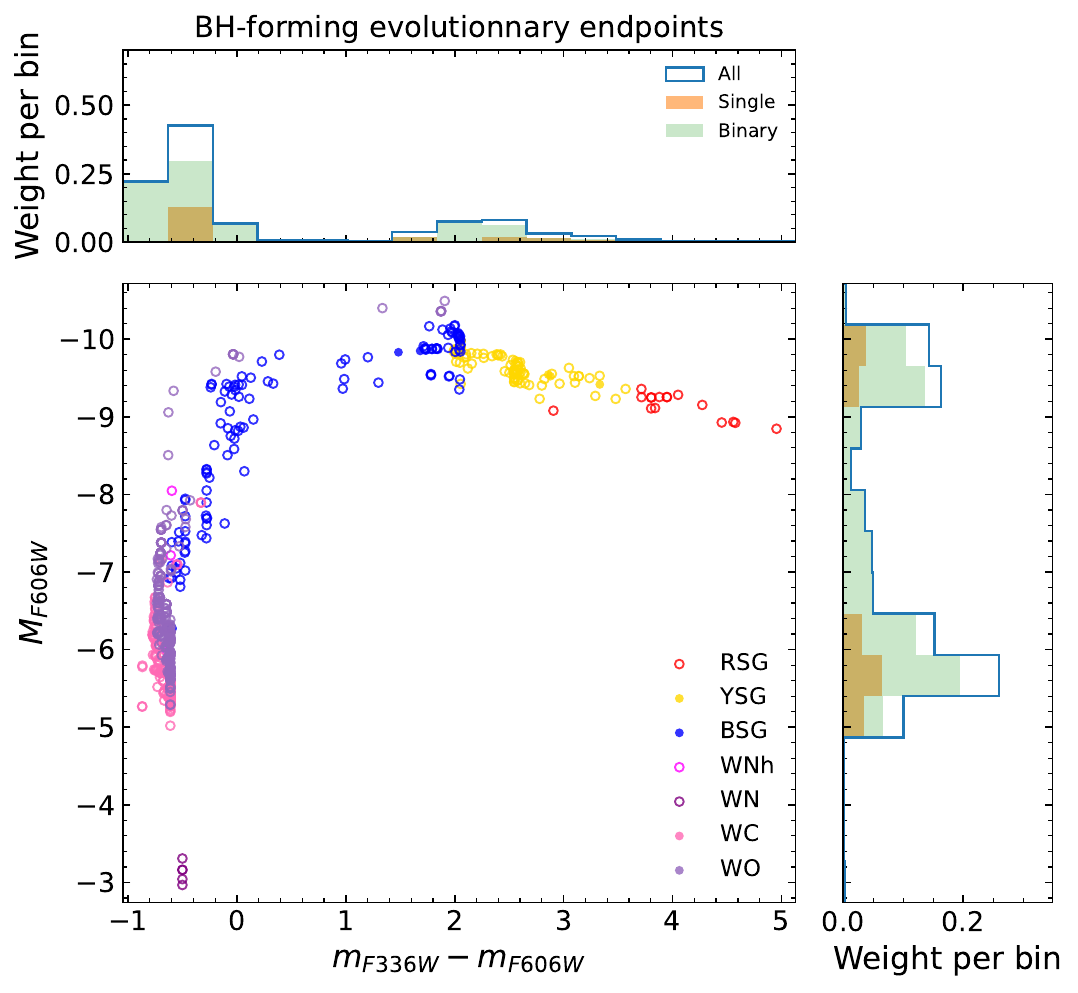}
        \caption{$m_\mathrm{F336W} - m_\mathrm{F606W}$ vs.\ $M_{\rm F606W}$}
    \end{subfigure}
    \begin{subfigure}[t]{0.49\textwidth}
        \centering
        \includegraphics[width=\textwidth]{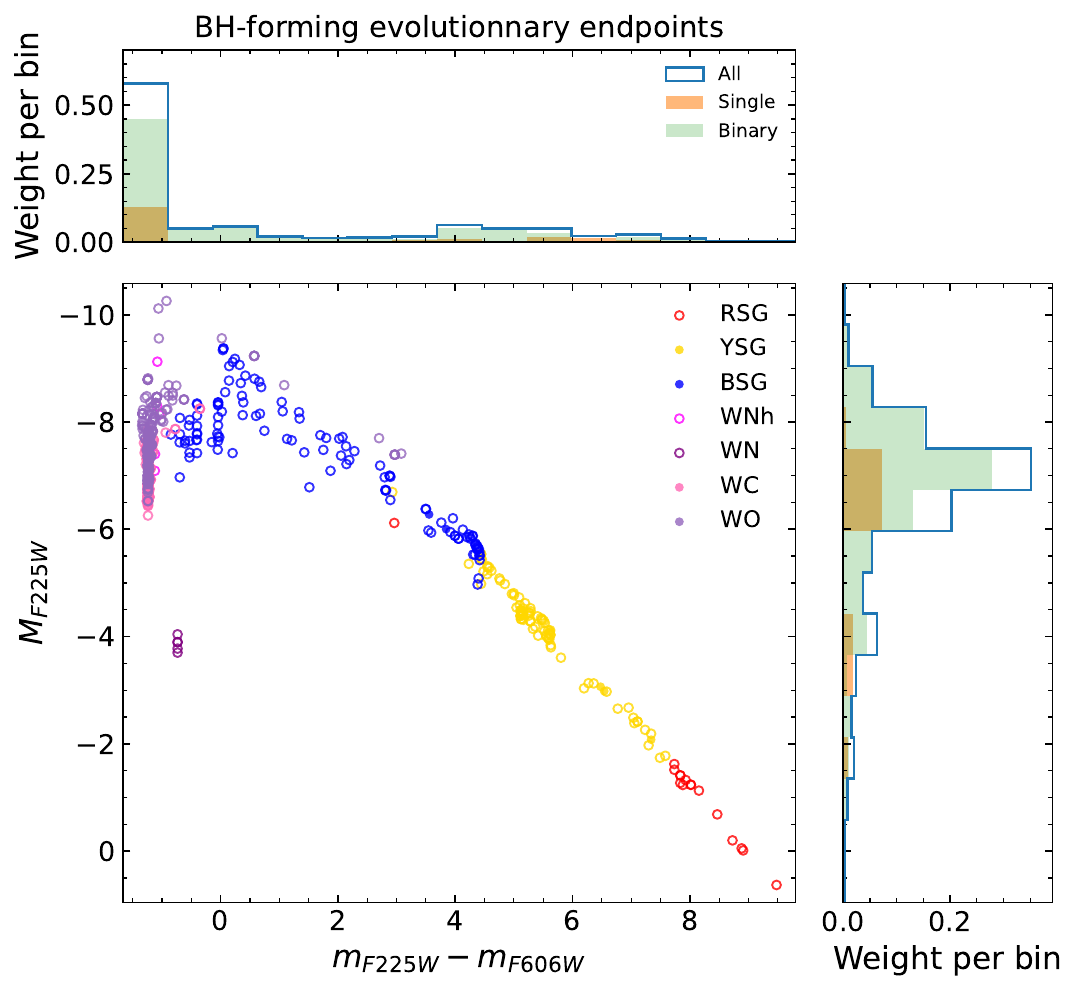}
        \caption{$m_\mathrm{F225W} - m_\mathrm{F606W}$ vs.\ $M_{\rm F225W}$}
    \end{subfigure}

    \caption{Colour--magnitude diagrams for BH-forming progenitors at $Z=0.014$, including all endpoints (single stars and binary systems). The plotted photometry includes the flux contribution of any non-degenerate companion present at collapse, while the colour coding/classification of each point is based on the properties of the BH-forming progenitor itself. The panels span optical/near-IR (F606W and F814W) to near-/far-UV (F336W and F225W) colours. The colour histograms on the side and absolute magnitude histograms above each panel show weighted distributions, accounting for the IMF, assumed binary fraction ($70\%$) and distributions of initial orbital configurations.}
    \label{fig:cmd_all}
\end{figure*}
Using the synthetic photometry and statistical weights described above, we construct probability-weighted colour--magnitude distributions of core-collapse progenitors. In Fig.\,\ref{fig:cmd_CCSNe_plus_BHs} we present all endpoints leading either to a CCSN or the formation of a BH in F336W (near-UV) and F606W (broad-band V) \textit{Hubble Space Telescope} (\textit{HST}) filters. CCSN progenitors are divided into subtypes according to their pre-collapse hydrogen and helium envelope masses.\footnote{We classify progenitors as Type~II-P for total hydrogen mass $M_{\mathrm{H}}>0.5\,\mathrm{M}_\odot$, Type~IIb for $0.033<M_{\mathrm{H}}/\mathrm{M}_\odot\leq0.5$, Type~Ib for $M_{\mathrm{H}}\leq0.033\,\mathrm{M}_\odot$ and helium mass $M_{\mathrm{He}}>0.5\,\mathrm{M}_\odot$, and Type~Ic otherwise, following \citetalias{Gilkis2025}.} The contribution of a non-degenerate companion (if present) to the flux is included, though we note that many binary evolution endpoints do not have such contributions because the primary star has already ended its life, or the two stars merged. This contribution can shift some otherwise cool progenitors to bluer apparent colours; for example, a subset of binary Type~II-P progenitors appear bluer than the single-star RSG locus in Fig.\,\ref{fig:cmd_CCSNe_plus_BHs} because of flux from their companions. The CCSN and BH-forming populations are distinct, with CCSN progenitors occupying a wide range of colour--magnitude space in these chosen filters, but BH progenitors clustering around blue and UV-bright regions.
All distributions are shown in terms of intrinsic (dust-free) absolute AB magnitudes.

In Fig.\,\ref{fig:cmd_all} we focus only on BH progenitors, and include also the F225W (far-UV) and F814W (red/near-IR) filters, including the contribution from a non-degenerate companion if present. We note several features that are apparent in Fig.\,\ref{fig:cmd_all}. In all the colour indices that we compute, the weighted distribution shows an overwhelming tendency for BH progenitors to blue colours. This is most notable in the panel showing $m_\mathrm{F225W} - m_\mathrm{F606W}$ vs.\ $M_{\rm F225W}$, where most progenitors are both blue and very bright in the far-UV filter. While a non-negligible fraction are rather bright in all filters, the majority of progenitors are significantly fainter in visible range filters compared to near-UV and far-UV, stressing the importance of observations in UV to identify stars disappearing when forming BHs. Potential modifications to the apparent progenitor classifications arising from late-time LBV-like excursions are shown in Appendix\,\ref{app:lbv}.

The same conclusion holds when considering only single-star progenitors (Appendix\,\ref{app:single}), indicating that the predominance of hot, blue BH progenitors is not solely a consequence of binary evolution. Appendix\,\ref{app:lowZ} presents the corresponding binary-inclusive colour--magnitude distributions at lower metallicities ($Z=0.0056$ and $0.00224$, approximately LMC/SMC). The LMC-metallicity results are qualitatively similar to the fiducial MW-metallicity case, while the SMC-metallicity models show a modest shift toward cooler BH progenitors but retain a substantial population of hot, blue systems.

\section{Prospects for future searches}
\label{sec:TheFutureIsBrightBlue}

The photometric predictions in Section~\ref{sec:TheyAreBlue} imply that the stellar populations most likely to undergo direct-collapse BH formation occupy a distinct region of colour--magnitude space: they are predominantly hot and blue, and they are especially luminous at short wavelengths. This has two immediate consequences. First, the optimal search strategy for disappearing stars should include blue/UV filters rather than relying solely on red-optical imaging that is tuned to cool supergiants. Second, interpreting candidate disappearances requires accounting for contamination by companions and for non-terminal phenomena such as dust formation and coalescing stars (Section\,\ref{sec:FailedSearches}), motivating a multi-band, high-resolution approach. In our fiducial population calculation, approximately one third of BH progenitors retain a luminous non-degenerate companion at the time of collapse, implying that companion light may affect the observed photometric properties and post-disappearance signatures of some systems.

Dust extinction may further complicate this interpretation. Because the colour--magnitude predictions presented here are intrinsic and dust-free, circumstellar or local extinction can shift BH progenitors to fainter and redder observed magnitudes. For example, modest extinction of $A_V\approx 1$\,mag would redden $m_\mathrm{F225W}-m_\mathrm{F606W}$ by roughly $1.7$\,mag while dimming $M_\mathrm{F606W}$ by nearly $1$\,mag, potentially moving intrinsically hot progenitors closer to the loci of cooler stars in observed colour--magnitude diagrams. Multi-band photometry is therefore important for distinguishing reddened hot progenitors from genuinely cool supergiants.

\subsection{Progenitor-type distribution}
\label{subsec:typedist}

Across all evolutionary channels considered here, we find that BH-forming progenitors are dominated by compact, stripped stars. The majority are WR stars ($\approx 60 \%$), with WC and WO sub-types contributing most strongly. A significant minority of BH progenitors are BSGs or YSGs ($\approx 38 \%$), arising from channels where substantial hydrogen remains at collapse but the core is sufficiently compact to fail explosion. RSGs constitute only a small fraction of BH-forming progenitors in our framework ($\approx 2 \%$).

This hierarchy is consistent with the colour--magnitude distributions: WR progenitors populate the bluest loci and are the most UV-bright, while BSG/YSG progenitors extend the distribution to somewhat redder colours. In contrast, the region occupied by luminous RSGs contains relatively few BH-forming endpoints, implying that searches focused exclusively on luminous cool stars test only a subset of BH-formation channels. As previously identified disappearing-star candidates were RSGs, we also provide a comparison between our endpoint synthetic photometry and the observational data of these suggested progenitors in Appendix\,\ref{app:disphot}. We also note that the predicted magnitudes of WR progenitors are broadly consistent with observed WR populations. Using the catalogue of LMC WR stars from \cite{Massey2002}, \cite{Eldridge2013} found absolute magnitudes in \textit{HST} optical filters spanning approximately $-2$ to $-8\,\mathrm{mag}$, comparable to the range predicted by our models. This agreement provides an observational consistency check for the adopted atmosphere mapping and synthetic photometry.

\subsection{Event rates}
\label{subsec:rates}

Using the population weights described in Section~\ref{subsec:weights}, we compute the expected rate of BH formation without a successful supernova per unit star-formation rate, and we break the total into contributions by pre-collapse stellar type. Table\,\ref{tab:BHratesByType} lists the resulting rates\footnote{The effect of reclassifying progenitors that may undergo late-time LBV-like phases is explored separately in Appendix\,\ref{app:lbvbh}.}, expressed as the number of disappearing-star events per century for a galaxy forming stars at $1\,\mathrm{M}_\odot\,\mathrm{yr}^{-1}$. The predicted rates depend on the adopted mapping between pre-collapse structure and explosion outcome, and we explore this uncertainty in Appendix\,\ref{app:explErtl} by repeating the calculation with the independent two-parameter explodability criterion of \citet{Ertl2016}. This increases the total BH formation rate at $Z=0.014$ from $0.418$ to $0.694$ per century, for the same assumed star-formation rate, primarily through an increased contribution from supergiants (particularly RSGs). Nevertheless, blue, hot WR and BSG progenitors still account for approximately $60$ per cent of BH formation. Thus, while the absolute rate and relative contributions of different progenitor types are sensitive to the core-collapse prescription, our qualitative prediction that most BH progenitors are hot and blue remains robust.

\begin{table}
\centering
\caption{Predicted rate of BH formation without a successful supernova (``disappearing stars'') per unit star-formation rate, by progenitor type, for $Z=0.014$. Rates are given in events per century for ${\rm SFR}=1\,\mathrm{M}_\odot\,\mathrm{yr}^{-1}$.}
\label{tab:BHratesByType}
\begin{tabular}{cc}
\hline
Stellar type & Rate [century$^{-1}$ $(M_\odot\,{\rm yr^{-1}})^{-1}$] \\
\hline
RSG  & 0.009 \\
YSG  & 0.068 \\
BSG  & 0.089 \\
HeG  & 0.000 \\
WNh  & 0.003 \\
WN   & 0.001 \\
WC   & 0.137 \\
WO   & 0.111 \\
\hline
Any & 0.418 \\
\hline
\end{tabular}
\end{table}

For context, the expected CCSN rate in the nearby Universe can be estimated empirically from local volumetric measurements obtained by untargeted transient surveys. A recent All-Sky Automated Survey for Supernovae study finds volumetric rates for CCSNe and their subtypes based on a well-controlled sample of nearby events, providing a total volumetric CCSN rate of $r_\mathrm{CCSN}\simeq 7\times 10^{-5}\, \mathrm{yr}^{-1}\,\mathrm{Mpc}^{-3}$ \citep{Pessi2025}. Within a sphere of radius $30\,\mathrm{Mpc}$, corresponding to a volume $V\approx 1.13\times 10^{5}\,\mathrm{Mpc}^{3}$, this implies an expected discovery rate of about $8$ CCSNe per year. The true rate may be modestly higher if a fraction of heavily obscured events are missed by optical searches. Adopting this empirical CCSN baseline, and using our model prediction of
around $3.7$ 
CCSNe per direct-collapse BH formation event, we expect a disappearing-star rate of approximately $1$ to $2\,\mathrm{yr}^{-1}$ within 30~Mpc. Restricting to a sample of around $100$ luminous, star-forming galaxies (which plausibly host a substantial fraction of the local star-formation budget), this corresponds to approximately one disappearing-star event per year, motivating decade-scale, multi-band monitoring with high-resolution UV/optical imaging.

The total predicted rate of disappearing stars is dominated by WC and WO progenitors, reflecting the high weights of stripped channels that produce compact cores and strong winds. BSG and YSG progenitors provide a substantial secondary contribution. The predicted contribution from RSG progenitors is small in comparison, and helium giants are negligible in our grid. These relative contributions motivate search strategies that prioritise sensitivity to hot, blue stars rather than focusing on the most luminous cool supergiants.

This estimate is also consistent with the results of the long-running LBT search for failed supernovae. The LBT survey monitored $27$ nearby galaxies out to approximately $10\,\mathrm{Mpc}$ over a baseline of about $11\,\mathrm{yr}$ \citep{Neustadt2021}. Importantly for the progenitor population predicted here, the survey included $U$-band imaging and was designed to identify persistent luminosity changes of order $|\Delta \nu L_\nu| \gtrsim 10^4\,\mathrm{L}_\odot$ \citep{Gerke2015,Adams2017_constraints,Neustadt2021}. This sensitivity encompasses a substantial part of the hot, blue progenitor population predicted here, with the bluer LBT bands providing greater sensitivity to such objects than the redder optical bands. However, the optically faintest stripped and WR progenitors may remain difficult to detect. The small number of plausible disappearing-star candidates identified over its $11\,\mathrm{yr}$ baseline is therefore broadly consistent with our predicted event rate once the substantially smaller volume and galaxy sample compared with our $30\,\mathrm{Mpc}$ estimate are taken into account. Extinction and crowding will further reduce the effective sensitivity to some progenitors.

\subsection{Implications for observational searches for black hole formation}
\label{subsec:hst}

The key observational requirement for detecting disappearing hot progenitors is deep, high-resolution imaging in blue/UV bands over a sufficiently large galaxy sample, followed by a long time baseline to identify the absence of previously detected sources. 

\begin{figure*}
    \centering
    \includegraphics[width=\textwidth]{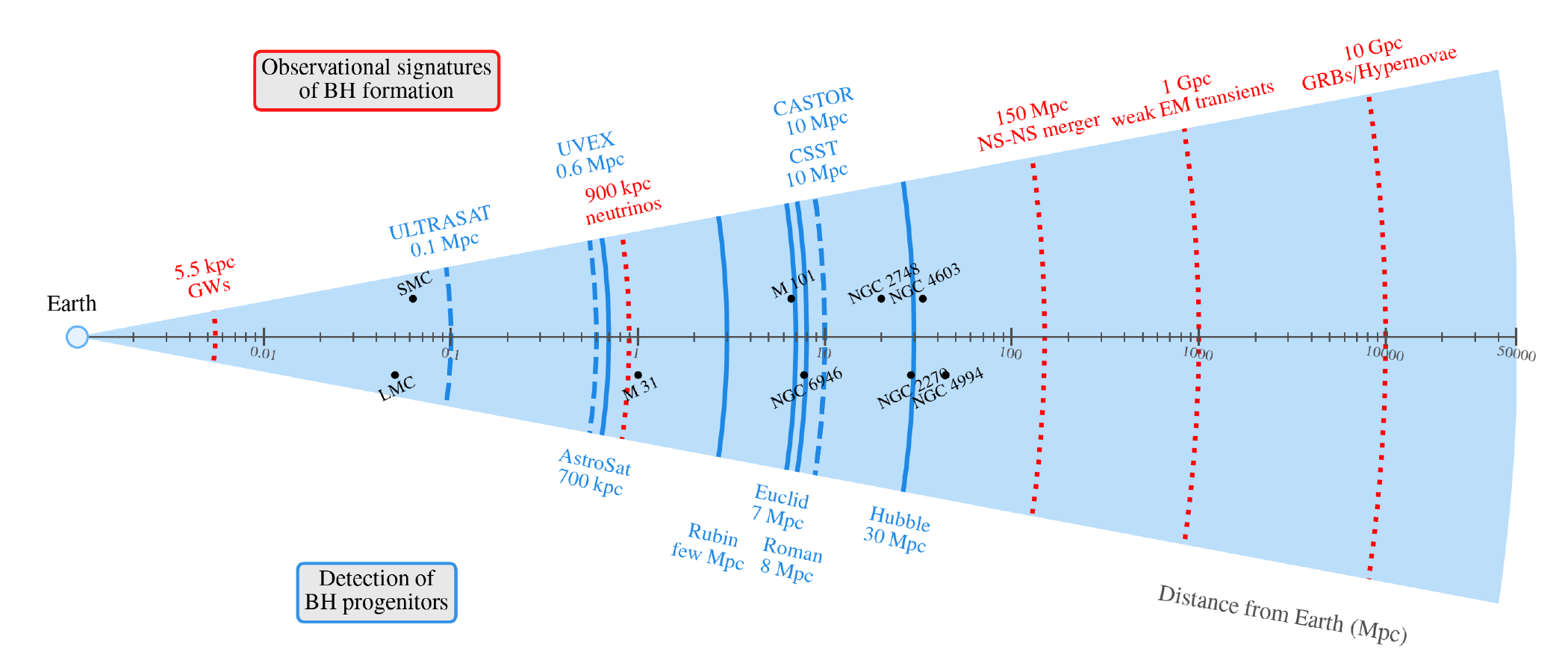}
    \caption{Overview of possible observational signatures of forming BHs and their progenitors in the nearby Universe as a function of their distance from Earth. Red dotted lines indicate the approximate limiting distance up to which different observational signatures can be found. Blue solid lines indicate the approximate limiting distance for finding the stellar progenitors of BHs with different telescopes, based primarily on their angular resolution. Dashed lines indicate future missions. For context, we show the distance to several galaxies for which progenitors of supernovae and potential BH progenitors have been detected with the \textit{Hubble Space Telescope}, and also show NGC~4994, the host galaxy of the merging double neutron star GW~170817 (and its accompanying kilonova).}
    \label{fig:finding-BHs}
\end{figure*}

We summarise the theoretically expected signatures of BH formation, along with estimates for the maximum distance at which they could be detected, in Fig.\,\ref{fig:finding-BHs}. Explosive and electromagnetically bright events accompanying BH formation, such as GRBs, may be observed at very large distances, reaching Gpc scales \citep[e.g.][]{Levan2025}. The possible transients accompanying a failed supernova explosion may be observed up to several Mpc, while other messengers, such as neutrinos and GW signatures, are limited to nearby events \citep{Janka2012,Valtonen-Mattila2023ApJ,IceCube2024ApJ,Mueller2026GWreview}. In the bottom part of Fig.\,\ref{fig:finding-BHs}, we characterise the approximate maximum distances at which ongoing and proposed missions could detect BH progenitors, based on our findings and their angular resolution, which is key to identifying progenitors among crowded regions. We employ an approximate distance-scaling scheme, with the capabilities of HST serving as a baseline for estimating crowding effects. Taking the angular resolution for HST as ~0.05'' and the furthest detected and resolved UV-bright SN progenitors with HST at about $33\,\mathrm{Mpc}$, we infer the farthest distance for the other telescopes to resolve BH progenitors as $d = 33 \cdot \frac{0.05''}{\theta}\rm{\,Mpc}$, where $\theta$ is their angular resolution. 

As shown in Fig.\,\ref{fig:finding-BHs}, \textit{HST} is uniquely well suited to this task because it combines (i) spatial resolution required to deblend crowded massive-star fields in nearby galaxies, and (ii) access to UV/near-UV filters that are closer to the SED peaks of hot stars. UV imaging is particularly valuable because the dominant progenitor classes (WC/WO and many BSGs) are intrinsically UV-bright and occupy cleaner loci in UV–optical colour–magnitude space, improving discrimination from cooler stellar populations and reddened contaminants.

A practical strategy is therefore a multi-epoch imaging programme of a large, well-defined sample of nearby star-forming galaxies with high angular resolution instruments. A minimum of two epochs is required to identify vanished sources, though three or more epochs are preferable to establish that candidate progenitors were persistent prior to disappearance rather than transient or highly variable contaminants. The initial imaging should obtain near-UV and optical data (e.g.\ WFC3/UVIS with filters such as F225W/F336W together with optical bands such as F606W and F814W), reaching depths sufficient to detect the relevant progenitor populations across the full distance range of interest. A later epoch, obtained after a baseline of order a decade, then enables a direct search for vanished sources via image subtraction and/or catalogue-level cross-matching\footnote{Existing long-baseline monitoring programmes such as the Large Binocular Telescope survey \citep{Kochanek2008, Adams2017, Adams2017_constraints, Neustadt2021} already provide deep optical imaging for a substantial nearby galaxy sample and therefore remain highly valuable for constraining disappearing-star populations. However, sensitivity to the hottest stripped progenitors may still benefit from additional ultraviolet coverage and high angular resolution imaging.}.

This approach naturally complements ongoing and future wide-field time-domain surveys. Facilities such as \textit{Rubin} \citep{Ivezic2019}, \textit{Euclid} \citep{Inserra2018}, \textit{Roman} \citep{Hounsell2018}, and \textit{JWST} \citep{DeCoursey2025} will deliver complementary wide-field monitoring in optical/IR bands, but they do not provide the same combination of UV coverage and sub-arcsecond resolution for crowded fields that is needed to robustly identify hot progenitors and disentangle companions. Other ground-based time-domain surveys with wide fields, such as the Zwicky Transient Facility, provide additional coverage and have the potential to catch transients associated with black hole formation. Among these, \textit{BlackGEM} \citep{Groot2024BlackGEM} provides the best ground-based near-UV coverage based on its survey strategy, which includes a Sloan \textit{u} filter.  An \textit{HST}-anchored first epoch still provides an essential UV/optical baseline for decade-scale monitoring of failed supernova candidates, and directly targets the progenitor classes that dominate our predicted BH-formation channels. Given the predicted prevalence of WR-type progenitors, the use of dedicated filters that can detect emission lines from such stars would be a valuable tool for confirming the disappearance of such progenitors.

We also highlight the potential of future proposed missions for undertaking this type of search. Wide-field ultraviolet survey missions such as \textit{ULTRASAT} \citep{Shvartzvald2024} and \textit{UVEX} \citep{Kulkarni2021} will provide broad coverage of the nearby Universe, though their relatively low angular resolution will limit their effectiveness in crowded star-forming regions. By contrast, \textit{CASTOR} \citep{Cote2012,Cote2025} and \textit{CSST} \citep{Cao2018} combine wide-field survey strategies with near-\textit{HST}-class angular resolution (around $0.1\arcsec$ to $0.15\arcsec$), making them particularly well suited to discovering BH progenitors within $10\,\mathrm{Mpc}$. We note that the proposed \textit{Lazuli} mission \citep{Roy2026}, with a similar angular resolution as \textit{HST} in the optical, will also have the potential to find BH progenitors up to about $30\,\mathrm{Mpc}$. However, the lack of a UV instrument and the operation strategy focused on rapid follow-up imply that it would be best used as a complementary follow-up imaging instrument. Detailed yield forecasts as a function of survey depth, cadence, and galaxy sample will be presented in future work.

\section{Discussion and summary}
\label{sec:TheseAreTheStarsYouShouldBeLookingFor}

We have presented predictions for the observable appearance of stars that are likely to undergo direct collapse to BHs without a visible supernova. By combining detailed stellar-evolution models with atmosphere libraries and population weighting, we derived colour–magnitude distributions and event-rate estimates across multiple evolutionary channels.

A key result is that most BH progenitors are predicted to be hot and blue, often in WR phases, and therefore most luminous at short wavelengths. This contrasts with the historical focus on red supergiants in failed-supernova searches. Our results suggest that broadening search strategies to include hot, UV-bright stars could significantly improve sensitivity to direct-collapse events.

The robustness of these conclusions depends on several modelling assumptions. Uncertainties in mass-loss rates, particularly in phases such as LBV evolution, may affect the relative fractions of stripped and hydrogen-rich progenitors, and thus the detailed distribution of colours (Appendix\,\ref{app:lbv} and Appendix\,\ref{app:lbvbh}). The absolute BH-formation rate and relative contributions of different progenitor types also depend on the adopted core-collapse prescription. However, the predominance of hot, blue progenitors persists when adopting an alternative explodability criterion (Appendix\,\ref{app:explErtl}).

Binary evolution broadens the diversity of progenitors and enhances the production of stripped stars, but we find that the dominance of hot, blue progenitors persists even when considering single-star evolution alone. This indicates that our main result is not driven solely by binary interactions.

The presence of a luminous companion may affect the observed colours and magnitudes of progenitor systems, potentially shifting them toward redder or brighter regions of colour–magnitude space. However, many endpoints either lack a significant companion contribution or are coalescence products, limiting the impact of this effect on the overall population. In addition, dust formation and circumstellar obscuration may further complicate the identification of disappearing stars, reinforcing the need for multi-wavelength observations (see also Appendix\,\ref{app:disphot}).

Our results imply that the most effective searches for disappearing BH progenitors require observational capabilities that differ substantially from those optimised for red-supergiant progenitor studies. In particular, robust identification of the predicted hot, blue BH-forming population demands sensitivity at near- and preferably far-ultraviolet wavelengths, together with sub-arcsecond angular resolution to mitigate crowding and disentangle companions in star-forming regions. High-resolution UV imaging with facilities such as \textit{HST}, and in the future potentially \textit{CASTOR}, would therefore provide particularly powerful constraints on disappearing-star populations, while wide-field optical/IR surveys from facilities such as \textit{Rubin}, \textit{Roman}, and \textit{Euclid} can offer complementary long-baseline monitoring and transient follow-up.

Future time-domain surveys and targeted monitoring campaigns will provide critical tests of our predictions. Continued synergy between stellar modelling and observational searches will be essential to understand the physics of core collapse and black hole formation.

\section*{Acknowledgments}

We thank an anonymous referee for a constructive review of the paper.
We thank Hugues Sana for helpful discussions.
AG and EL thank the Kavli Institute for Cosmology visitor programme that enabled a short visit at the Institute of Astronomy and brought about the discussion from which this paper emerged.
AG and CAT acknowledge support from the Isaac Newton Trust (University of Cambridge).
EL acknowledges funding through a start-up grant from the Internal Funds KU Leuven (STG/24/073), a Veni grant  (VI.Veni.232.205) from the Netherlands Organization for  Scientific Research (NWO), and the Research Foundation – Flanders (FWO) under the Odysseus Program, Type II (G0AT525N).
CDK gratefully acknowledges support from the NSF through AST-2432037, the HST Guest Observer Program through HST-SNAP-17070 and HST-GO-17706, and from JWST Archival Research through JWST-AR-6241 and JWST-AR-5441.
AO acknowledges support from the McWilliams Postdoctoral Fellowship.
CAT thanks Churchill College for his fellowship.

\section*{Data Availability Statement}

The data underlying this article can be found at \href{https://zenodo.org/records/22229107}{https://zenodo.org/records/22229107}.

\bibliographystyle{mnras}
\input{HBBH.bbl}

\appendix

\section{Potential Late-Time LBV-like Behaviour in Black Hole Progenitors}
\label{app:lbv}

Some BH-forming progenitors may pass through LBV-like phases shortly before collapse, potentially altering their observational appearance and the interpretation of disappearing-star searches. To assess this possibility, we repeat our classification analysis for the $Z=0.014$ population while identifying endpoints that may have experienced LBV-like behaviour during the final stages of evolution.

Specifically, we examine the final $10^5$\,yr prior to core collapse. We classify a progenitor as an LBV candidate if it spends at least $10^4$\,yr of this final interval within the LBV regime defined by the instability-strip criteria of \citet{Schneider2021,Schneider2024}. This threshold is motivated by the expectation that shorter excursions are unlikely to drive substantial cumulative LBV-like mass loss.

\begin{figure}
    \centering
    \includegraphics[width=\linewidth]{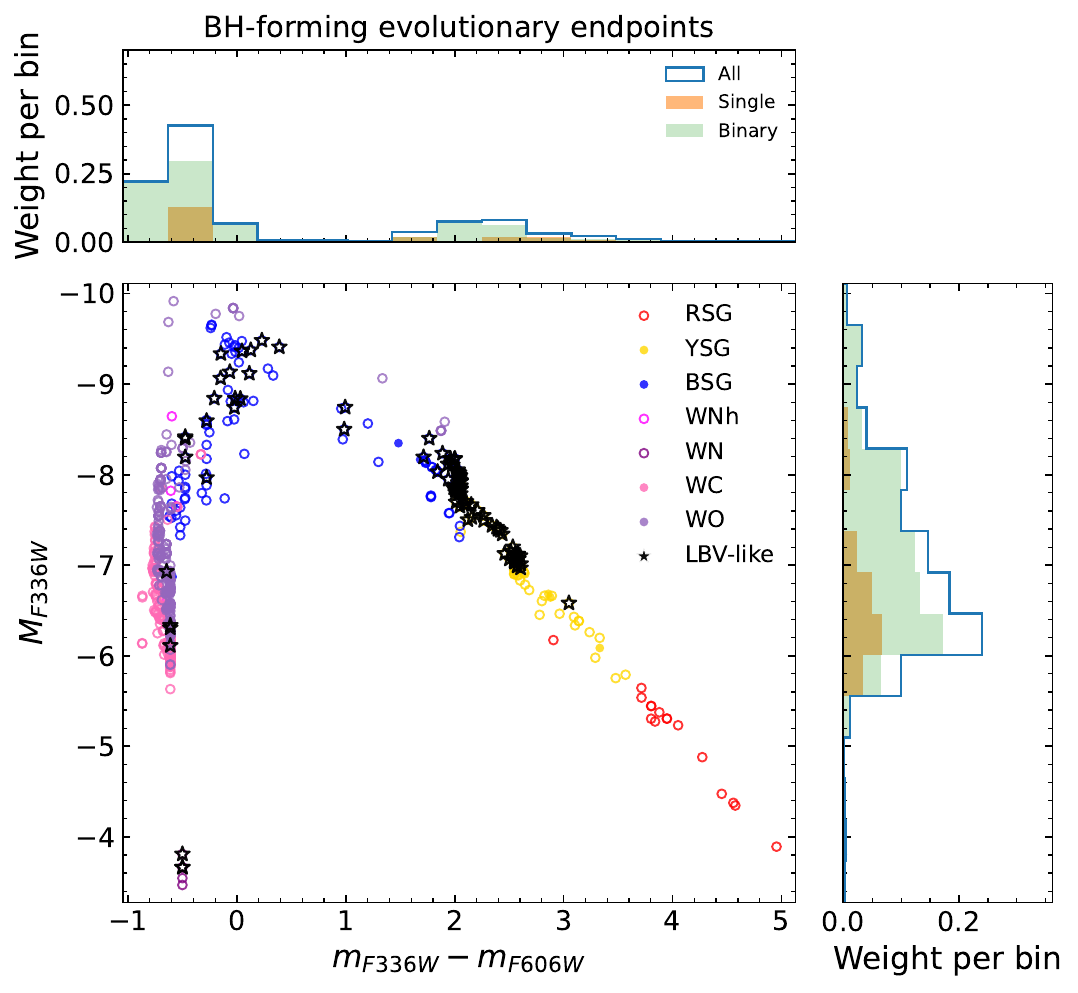}
    \caption{Colour--magnitude diagram for all endpoints predicted to produce either CCSNe or direct-collapse BHs at $Z=0.014$, including all endpoints (single stars and binary systems), like Fig.\,\ref{fig:cmd_all} but with progenitors that potentially had LBV-like phases marked with black stars.}
    \label{fig:cmd_lbv}
\end{figure}

Fig.\,\ref{fig:cmd_lbv} highlights BH progenitors satisfying this criterion in a colour--magnitude diagram. The LBV-like progenitors are concentrated primarily among systems otherwise classified as BSGs and YSGs, consistent with the overlap of the adopted LBV instability region with those temperature ranges. A smaller number arise from stripped WR progenitors whose preceding evolution includes brief excursions through LBV-like regions before reaching their final hot stripped states. These results suggest that some disappearing BH progenitors may observationally resemble LBVs shortly before collapse.

\section{Single-star black hole progenitor colour--magnitude diagrams}
\label{app:single}

\begin{figure*}
    \centering
    \begin{subfigure}[t]{0.49\textwidth}
        \centering
        \includegraphics[width=\textwidth]{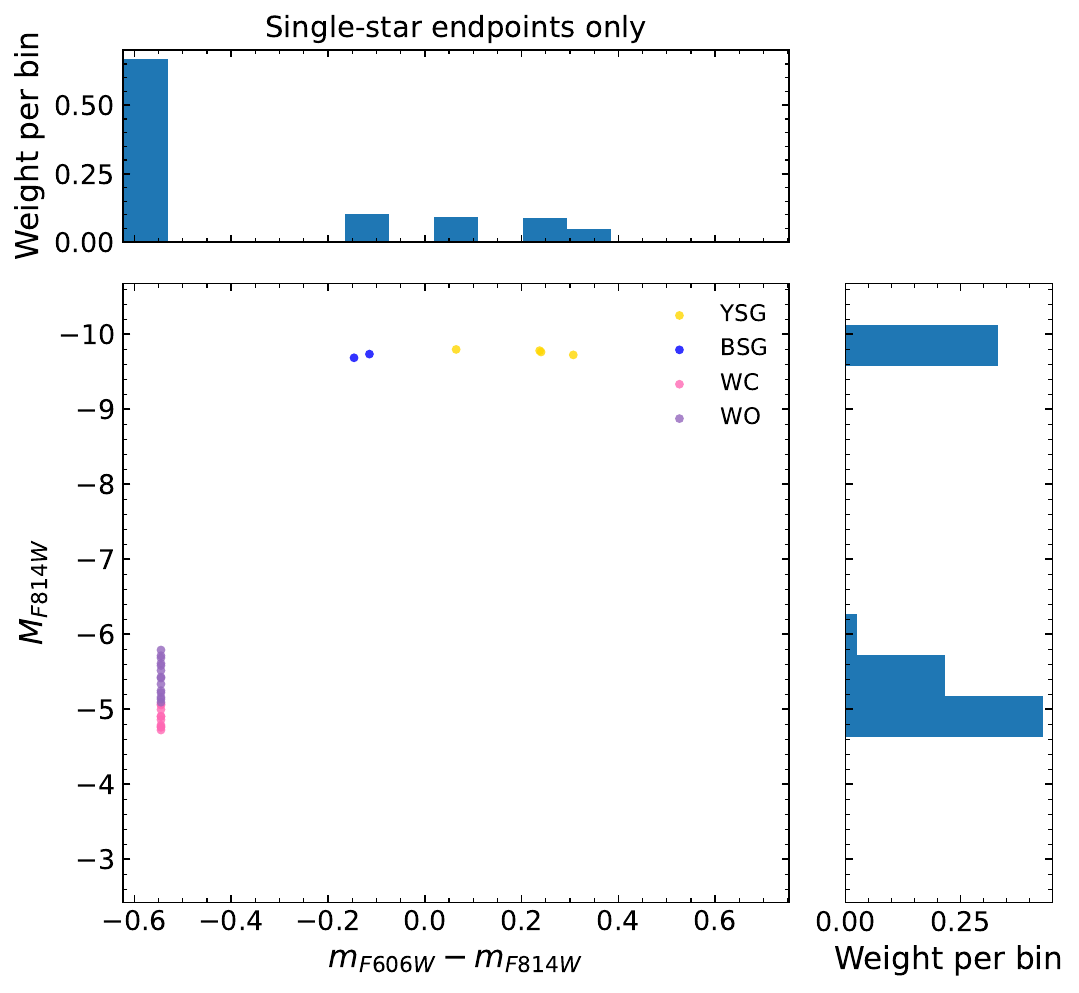}
        \caption{$m_\mathrm{F606W} - m_\mathrm{F814W}$ vs.\ $M_{\rm F814W}$}
    \end{subfigure}
    \begin{subfigure}[t]{0.49\textwidth}
        \centering
        \includegraphics[width=\textwidth]{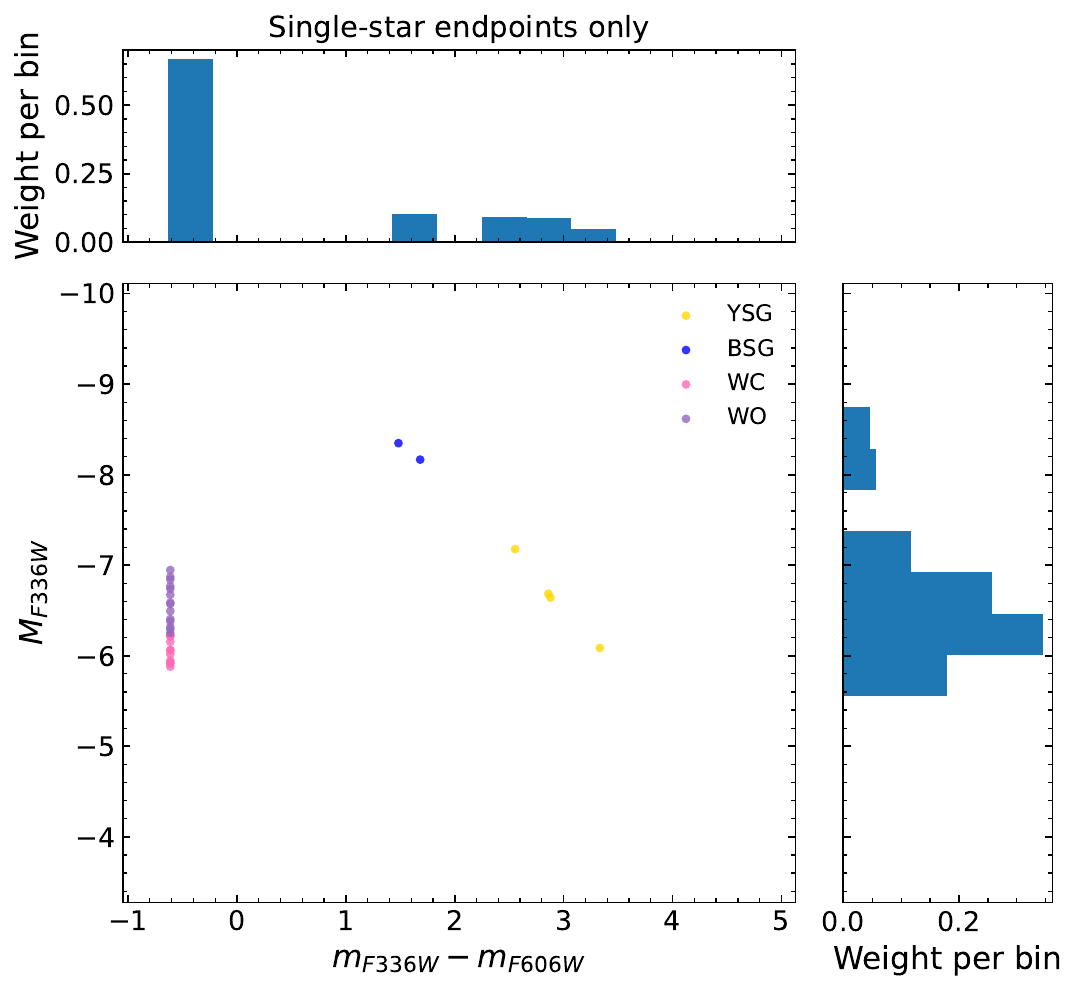}
        \caption{$m_\mathrm{F336W} - m_\mathrm{F606W}$ vs.\ $M_{\rm F336W}$}
    \end{subfigure}

    \vspace{0.8em}

    \begin{subfigure}[t]{0.49\textwidth}
        \centering
        \includegraphics[width=\textwidth]{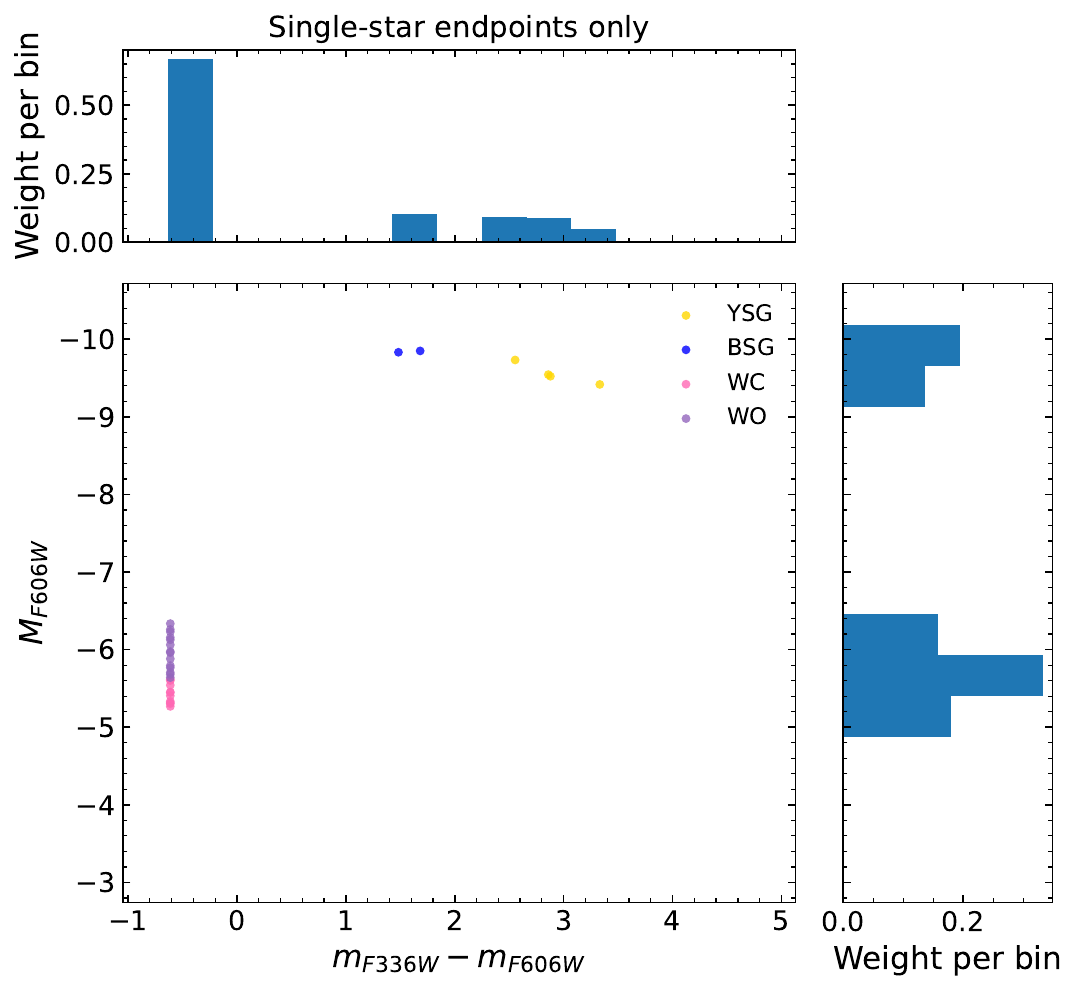}
        \caption{$m_\mathrm{F336W} - m_\mathrm{F606W}$ vs.\ $M_{\rm F606W}$}
    \end{subfigure}
    \begin{subfigure}[t]{0.49\textwidth}
        \centering
        \includegraphics[width=\textwidth]{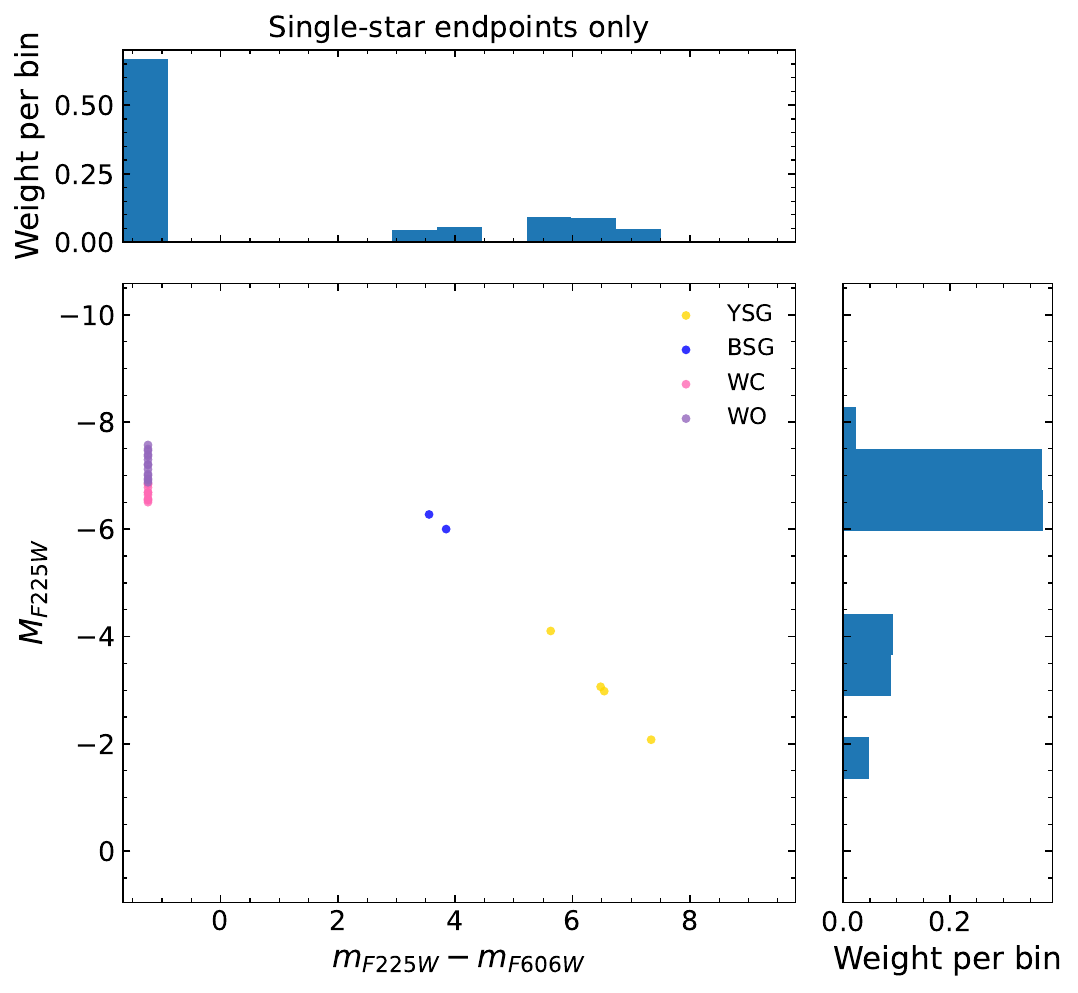}
        \caption{$m_\mathrm{F225W} - m_\mathrm{F606W}$ vs.\ $M_{\rm F225W}$}
    \end{subfigure}

    \caption{Same as Fig.\,\ref{fig:cmd_all}, but restricted to single-star evolution. Axis limits match those of Fig.\,\ref{fig:cmd_all}.}
    \label{fig:cmd_single}
\end{figure*}
For completeness, Fig.\,\ref{fig:cmd_single} shows the predicted colour--magnitude distributions of BH-forming progenitors when considering only single-star evolutionary endpoints. Axis limits are kept identical to those of the binary-inclusive colour--magnitude diagrams (Fig.\,\ref{fig:cmd_all}). The overall result is qualitatively consistent with the full population: BH progenitors remain predominantly hot and blue, demonstrating that this conclusion is not driven solely by binary interactions. However, the diversity of endpoints is reduced when binaries are excluded. In particular, no single-star BH progenitors occupy the RSG region of the colour--magnitude diagrams in our models; instead, all remain comparatively compact and lie in yellow or blue regions of colour--magnitude space.

\section{Metallicity Dependence of Predicted Progenitor Distributions}
\label{app:lowZ}

\begin{figure}
    \centering

    \begin{subfigure}[t]{0.49\textwidth}
        \centering
        \includegraphics[width=\textwidth]{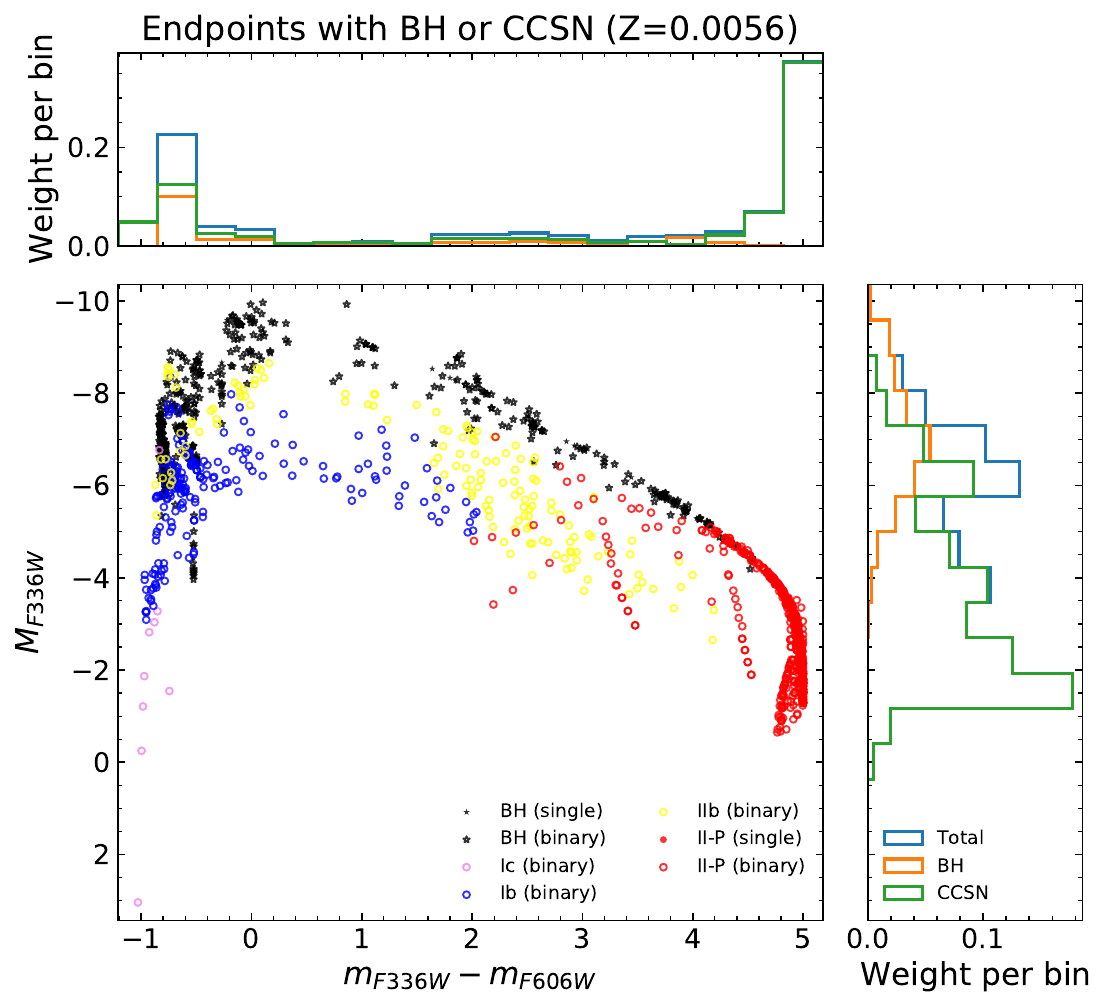}
        \caption{LMC metallicity ($Z = 0.0056$)}
    \end{subfigure}
    \begin{subfigure}[t]{0.49\textwidth}
        \centering
        \includegraphics[width=\textwidth]{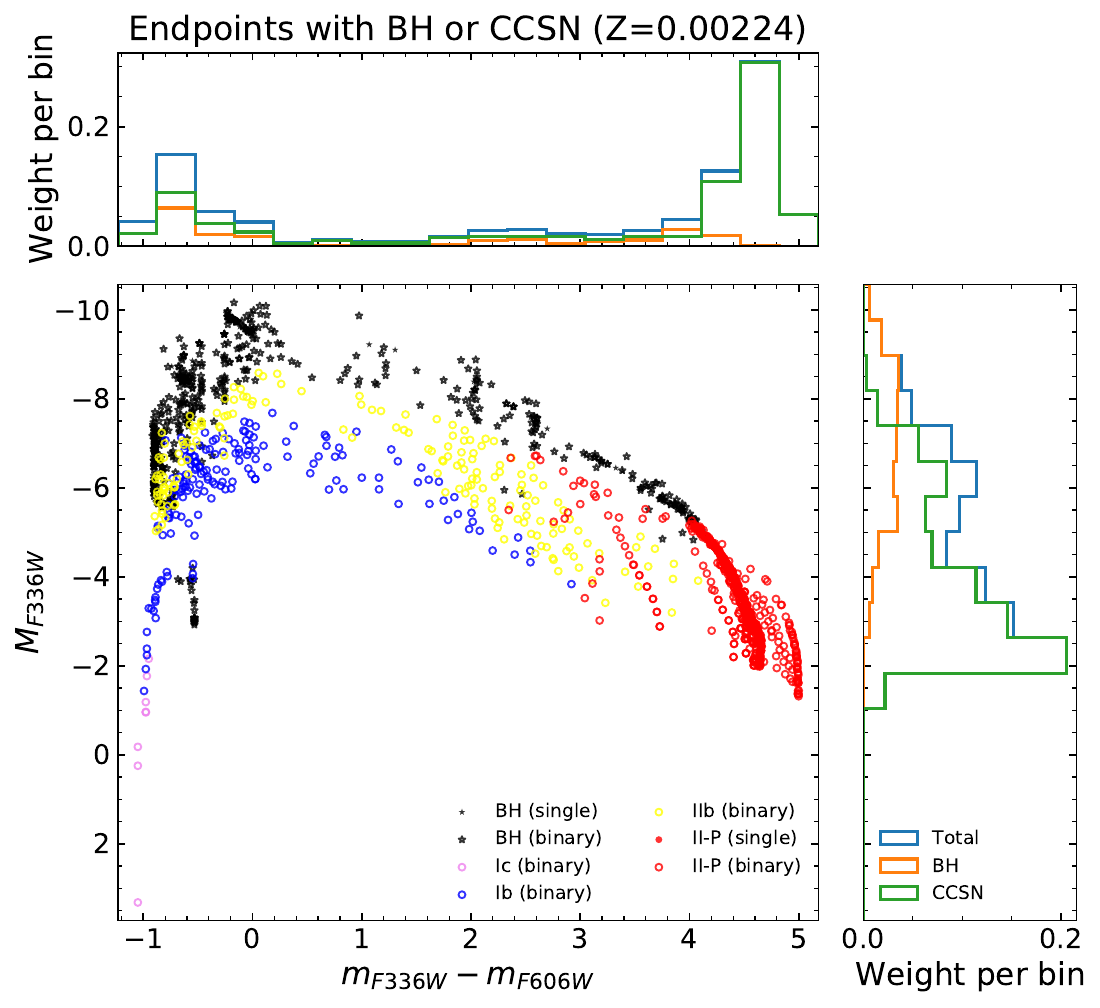}
        \caption{SMC metallicity ($Z = 0.00224$)}
    \end{subfigure}

    \caption{Same as Fig.\,\ref{fig:cmd_CCSNe_plus_BHs}, but for $Z=0.0056$ (top) and $Z=0.00224$ (bottom).}
    \label{fig:cmd_CCSNe_plus_BHs_lowZ}
\end{figure}

\begin{figure}
    \centering

    \begin{subfigure}[t]{0.49\textwidth}
        \centering
        \includegraphics[width=\textwidth]{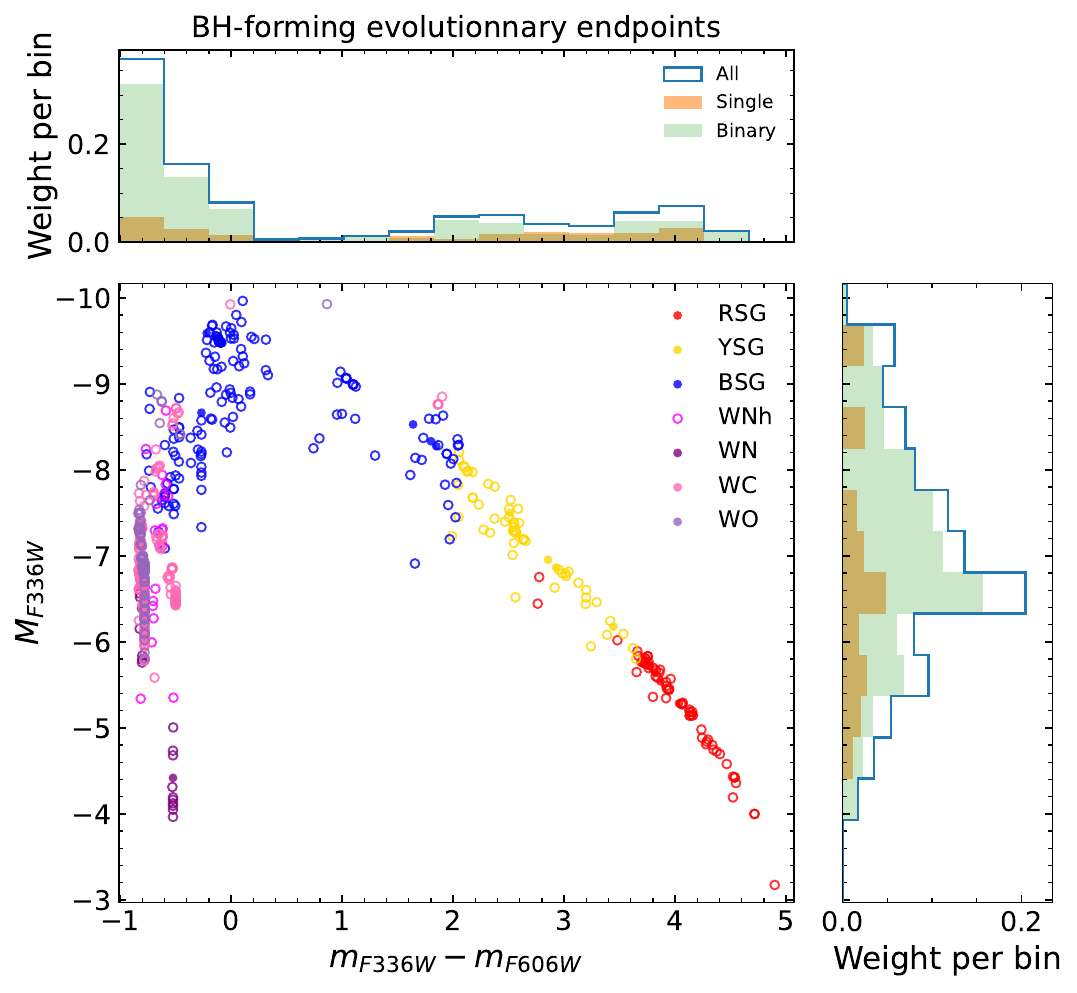}
        \caption{LMC metallicity ($Z = 0.0056$)}
    \end{subfigure}
    \begin{subfigure}[t]{0.49\textwidth}
        \centering
        \includegraphics[width=\textwidth]{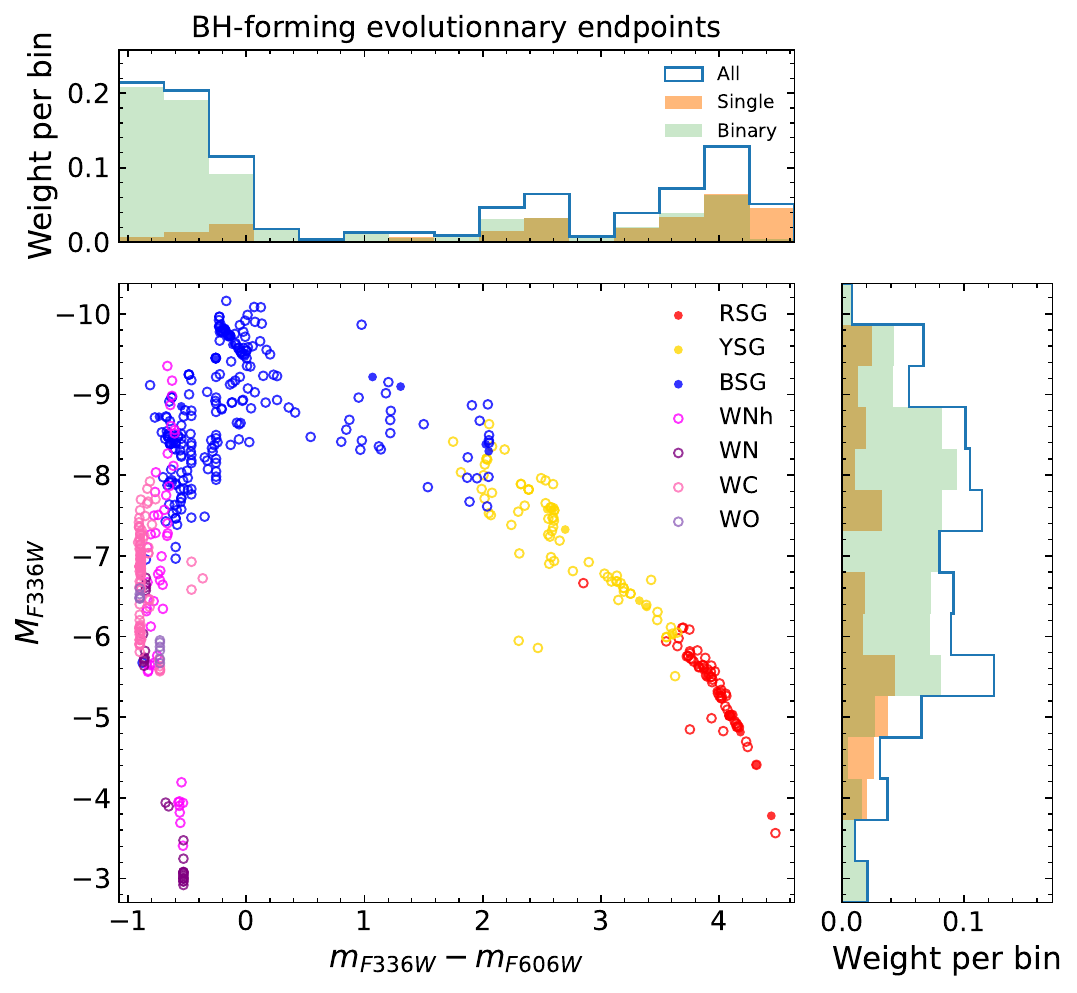}
        \caption{SMC metallicity ($Z = 0.00224$)}
    \end{subfigure}

    \caption{Colour--magnitude diagrams for BH-forming progenitors at $Z=0.0056$ (top) and $Z=0.00224$ (bottom), including all endpoints (single stars and binary systems), similar to Fig.\,\ref{fig:cmd_all}.}
    \label{fig:cmd_all_lowZ}
\end{figure}

To assess the sensitivity of our predictions to metallicity, we repeat the analysis of the fiducial population for $Z=0.0056$ and $0.00224$, approximately corresponding to LMC- and SMC-like metallicities. Fig.\,\ref{fig:cmd_CCSNe_plus_BHs_lowZ} shows the resulting colour--magnitude distributions of all CCSN and BH-forming progenitors in the $M_\mathrm{F336W}$ versus $m_\mathrm{F336W}-m_\mathrm{F606W}$ plane, while Fig.\,\ref{fig:cmd_all_lowZ} shows the corresponding distributions restricted to BH-forming progenitors.

The LMC-metallicity results are qualitatively very similar to the $Z=0.014$ case. BH progenitors remain predominantly hot, blue, and UV-bright, with the bulk of the population occupying regions of colour--magnitude space distinct from RSGs. At SMC metallicity, the BH-forming population shifts modestly toward cooler colours, reflecting weaker winds and a greater tendency for stars to retain more extended envelopes prior to collapse. Nevertheless, a substantial fraction of BH progenitors remain hot and blue even at the lowest metallicity we consider.

Table~\ref{tab:BHratesByType_lowZ} summarises the predicted BH-formation rates by progenitor type for all three metallicities. The total BH-formation rate is about the same for all metallicities ($\approx 0.4$\,century$^{-1}$ per $\mathrm{M}_\odot\,\mathrm{yr}^{-1}$), while the relative contribution of progenitor types evolves significantly. In particular, the fraction of BH-forming RSGs and BSGs increases toward low metallicity, whereas the contribution from stripped WR-like progenitors decreases, consistent with reduced wind stripping in metal-poor environments.

\begin{table}
\centering
\caption{Predicted rate of BH formation without a successful supernova per unit star-formation rate, by progenitor type and metallicity. Rates are given in events per century for ${\rm SFR}=1\,\mathrm{M}_\odot\,\mathrm{yr}^{-1}$.}
\label{tab:BHratesByType_lowZ}
\begin{tabular}{cccc}
\hline
Stellar type & $Z=0.014$ & $Z=0.0056$ & $Z=0.00224$ \\
\hline
RSG  & 0.009 & 0.062 & 0.098 \\
YSG  & 0.068 & 0.061 & 0.069 \\
BSG  & 0.089 & 0.099 & 0.142 \\
HeG  & 0.000 & 0.000 & 0.000 \\
WNh  & 0.003 & 0.015 & 0.044 \\
WN   & 0.001 & 0.019 & 0.021 \\
WC   & 0.137 & 0.113 & 0.050 \\
WO   & 0.111 & 0.030 & 0.004 \\
\hline
Any  & 0.418 & 0.399 & 0.426 \\
\hline
\end{tabular}
\end{table}

\section{Observed disappearing-star candidates on colour--magnitude diagrams}
\label{app:disphot}

\begin{figure}
    \centering
    \includegraphics[width=\linewidth]{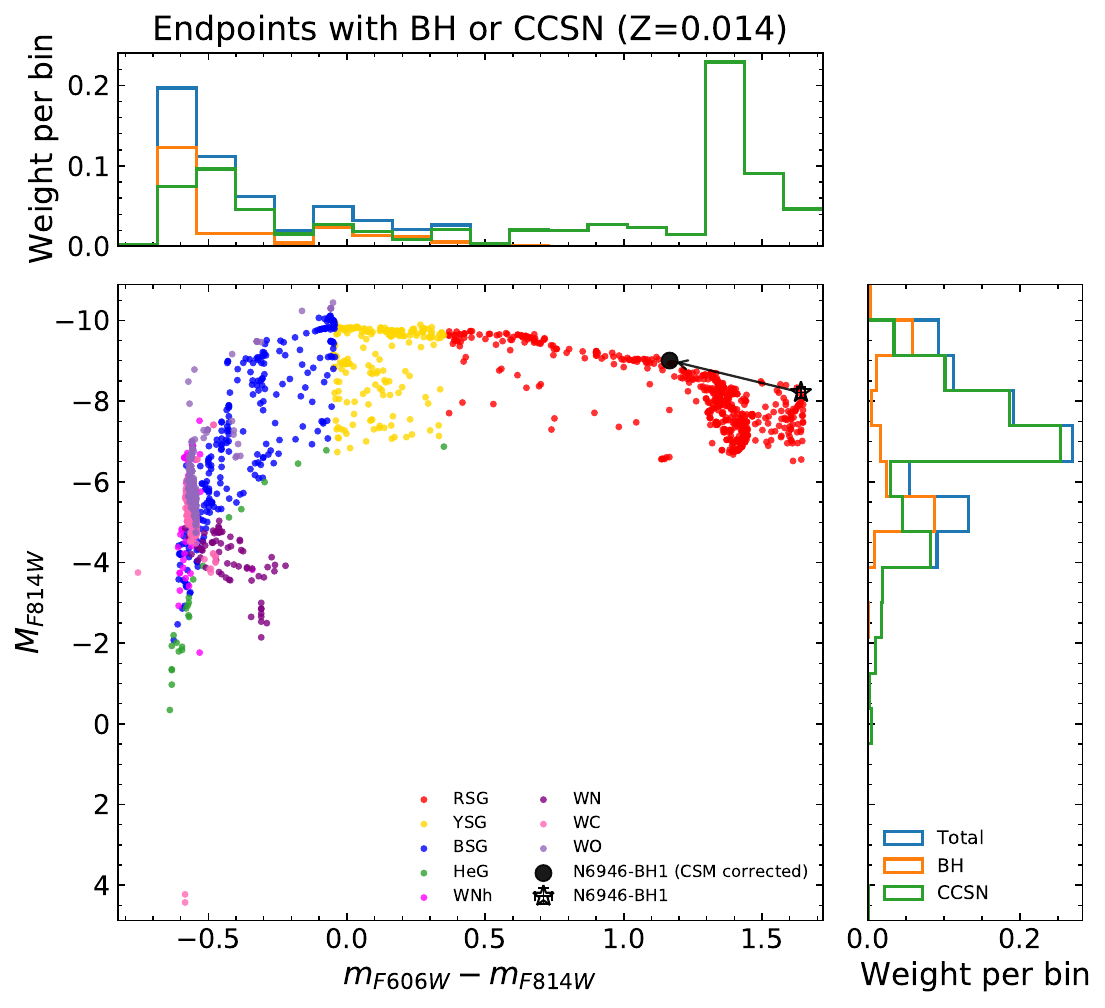}
    \caption{Colour--magnitude diagram for all endpoints predicted to produce either CCSNe or direct-collapse BHs at $Z=0.014$, including all endpoints (single stars and binary systems), compared to the observations of the disappearing-star candidate N6946-BH1.}
    \label{fig:cmd_N6946}
\end{figure}
\begin{figure}
    \centering
    \includegraphics[width=\linewidth]{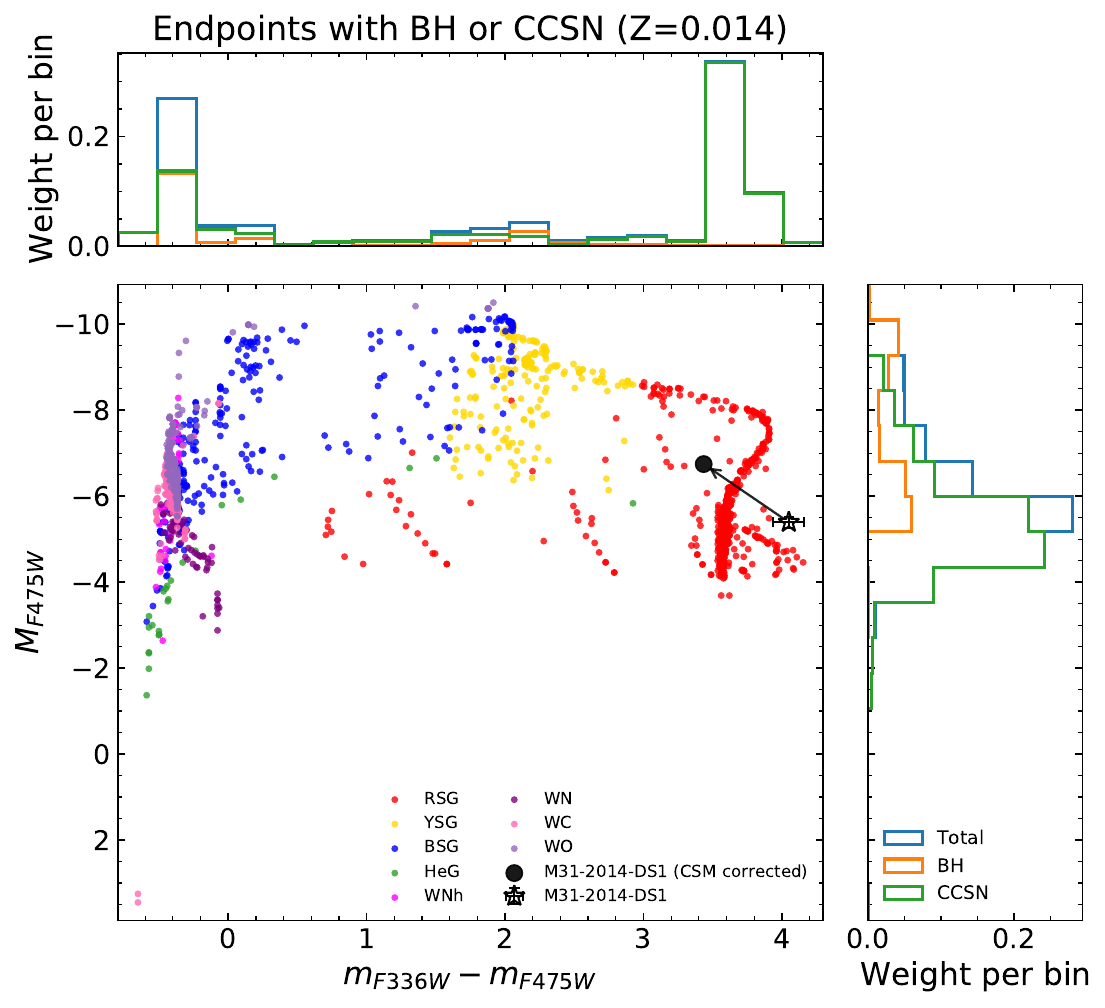}
    \caption{Like Fig.\,\ref{fig:cmd_N6946}, but with a comparison to the observations of the disappearing-star candidate M31-2014-DS1.}
    \label{fig:cmd_M31}
\end{figure}
We provide a comparison of our CMDs and observational data of two candidates for disappearing stars forming BHs: N6946-BH1 \citep{Adams2017} and M31-2014-DS1 \citep{De2026}. The observations are shown together with our model endpoints, marked by stellar type, in Fig.\,\ref{fig:cmd_N6946} (N6946-BH1) and Fig.\,\ref{fig:cmd_M31} (M31-2014-DS1).

For M31-2014-DS1, we adopt the pre-disappearance photometry from \citet{De2026}. The available HST photometry includes the filters F336W, F475W, F814W, F110W, and F160W. We adopt a distance to M31 of
$D = 770 \pm 40~{\rm kpc}$, following \citet{Savino2022}.

For N6946-BH1, we adopt the progenitor photometry quoted by \citet{Adams2017}, 
$F606W = 23.09 \pm 0.01$
and
$F814W = 20.77 \pm 0.01$,
originally identified in archival \textit{HST} imaging by \citet{Gerke2015}. These magnitudes were reported in the Vega system and were converted here to AB magnitudes.
We adopt a distance to NGC~6946 of
$D = 5.96 \pm 0.40~{\rm Mpc}$,
following \citet{Karachentsev2000}.

Foreground Galactic extinction corrections are applied to the observational points. For M31-2014-DS1 we adopt foreground corrections corresponding approximately to $A_V \simeq 0.1$, while for N6946-BH1 we adopt the Galactic extinction used by \citet{Adams2017}, corresponding to $E(B-V)=0.303$.

We also show approximate circumstellar-extinction-corrected photometry inferred from the modelling presented in \citet{Adams2017} and \citet{De2026}, indicated by arrows in Fig.\,\ref{fig:cmd_N6946} (N6946-BH1) and Fig.\,\ref{fig:cmd_M31} (M31-2014-DS1). For N6946-BH1, we adopt an effective circumstellar extinction of approximately $A_V \simeq 1.35$, while for M31-2014-DS1 we adopt approximately $A_V \simeq 1.23$, inferred from the reported optical depths and dust models. These corrected locations should be regarded as illustrative and model-dependent because both studies performed spectral energy distribution fitting using \texttt{DUSTY}.

The red colours of these two candidates occupy CMD regions where we do not predict many disappearing stars. One explanation for this is that our models under-predict the likelihood of RSGs collapsing to BHs, or that the two observed events are representatives of relatively rarer progenitors of BHs. Another possible explanation is that these events are not truly disappearing stars, as has been suggested in several studies and discussed in Section\,\ref{sec:FailedSearches}.

\section{Effect of LBV-like Reclassification on Predicted Event Rates}
\label{app:lbvbh}

To quantify the impact of possible late-time LBV-like phases on the predicted subtype distribution of BH-forming progenitors, we repeat the fiducial $Z=0.014$ event-rate calculation while reclassifying progenitors identified as LBV-like according to the criterion of Appendix\,\ref{app:lbv}.

The resulting BH-formation rates by apparent progenitor type are given in Table\,\ref{tab:BHratesByType_LBV}. We find that accounting for possible late-time LBV phases reclassifies a non-negligible fraction of BH progenitors as LBV-like, corresponding to a rate of $0.067$\,century$^{-1}$ per $\mathrm{M}_\odot\,\mathrm{yr}^{-1}$. This reclassification primarily reduces the inferred contributions of BSG and YSG progenitors, with smaller adjustments to the stripped WR categories. The total BH-formation rate is unchanged, but the apparent observational subtype distribution is modified accordingly.

\begin{table}
\centering
\caption{Predicted rate of BH formation without a successful supernova per unit star-formation rate at $Z=0.014$, reclassifying progenitors that spend at least $10^4$\,yr of their final $10^5$\,yr before collapse in the adopted LBV regime as LBV candidates.}
\label{tab:BHratesByType_LBV}
\begin{tabular}{cc}
\hline
Stellar Type & Rate [century$^{-1}$ $(M_\odot\,{\rm yr^{-1}})^{-1}$] \\
\hline
RSG  & 0.009 \\
YSG  & 0.030 \\
BSG  & 0.062 \\
HeG  & 0.000 \\
WNh  & 0.002 \\
WN   & 0.001 \\
WC   & 0.137 \\
WO   & 0.110 \\
LBV  & 0.067 \\
\hline
Any & 0.418 \\
\hline
\end{tabular}
\end{table}

\section{Sensitivity to explodability criteria}
\label{app:explErtl}

\begin{table}
\centering
\caption{Predicted rate of BH formation without a successful supernova per unit star-formation rate, by progenitor type, for $Z=0.014$, comparing our fiducial predictions with alternative results using the \citet{Ertl2016} criterion. Rates are given in events per century for ${\rm SFR}=1\,\mathrm{M}_\odot\,\mathrm{yr}^{-1}$.}
\label{tab:BHratesByTypeErtl}
\begin{tabular}{ccc}
\hline
Stellar type & Fiducial rate & Rate with Ertl criterion \\
\hline
RSG  & 0.009 & 0.190 \\
YSG  & 0.068 & 0.085 \\
BSG  & 0.089 & 0.104 \\
HeG  & 0.000 & 0.000 \\
WNh  & 0.003 & 0.011 \\
WN   & 0.001 & 0.046 \\
WC   & 0.137 & 0.149 \\
WO   & 0.111 & 0.110 \\
\hline
Any & 0.418 & 0.694 \\
\hline
\end{tabular}
\end{table}
As a test of the sensitivity of our results to the adopted core-collapse prescription, we repeat the classification of the progenitor models using the two-parameter explodability criterion of \citet{Ertl2016}. Specifically, the commonly used \texttt{s19.8} calibration of \citet{Ertl2016} to models by \citep{Woosley2002} is employed to separate successful core-collapse supernovae from failed explosions. Table\,\ref{tab:BHratesByTypeErtl} compares the resulting BH formation rates by progenitor type with those obtained using our fiducial \citet{Muller2016} prescription. The total predicted BH formation rate increases from $0.418$ to $0.694$ per century for a star-formation rate of $1\,\mathrm{M}_\odot\,\mathrm{yr}^{-1}$. The increase occurs across almost all progenitor classes, as the rate from WR stars increases from $0.25$ to $0.32$ per century. The only subtype showing a decrease is WO stars, for which the rate changes only marginally from $0.111$ to $0.110$ per century. Most of the overall increase is instead associated with supergiant progenitors, with the \citet{Ertl2016} criterion predicting substantially more BH formation from RSGs in particular. Helium giants do not contribute to BH formation under either prescription.

\begin{figure}
    \centering
    \includegraphics[width=\linewidth]{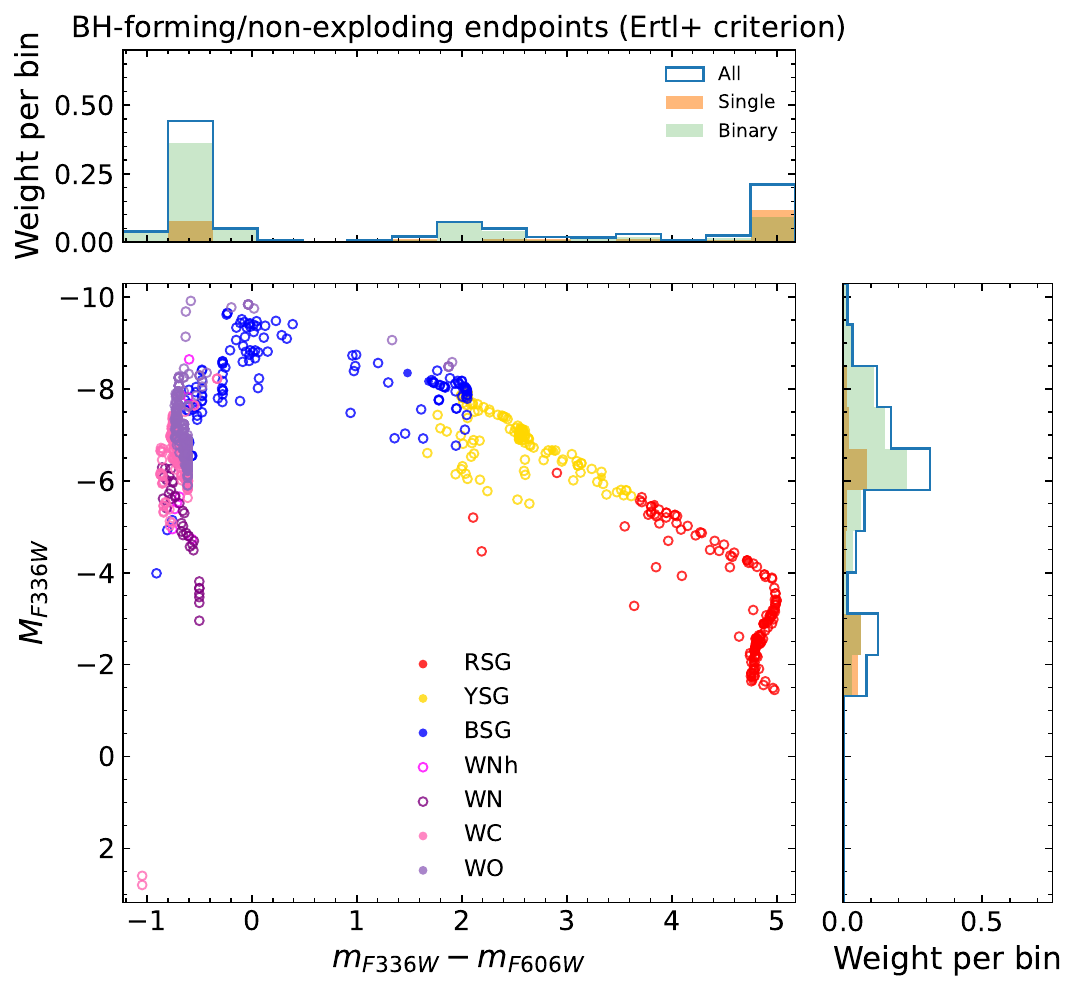}
    \caption{Colour--magnitude diagrams for BH-forming progenitors at $Z=0.014$, using the \citet{Ertl2016} explodability criterion. The two endpoints at the bottom-left should be considered non-exploding rather than BH-forming, as their final masses are below $2\,\mathrm{M}_\odot$.}
    \label{fig:cmd_Ertl}
\end{figure}
The corresponding progenitor distribution is shown in Fig.\,\ref{fig:cmd_Ertl} in the $m_\mathrm{F336W} - m_\mathrm{F606W}$ versus $M_{\rm F336W}$ plane. Compared with our fiducial results, the \citet{Ertl2016} criterion produces a substantially larger contribution from cool supergiants and reduces the relative dominance of WR progenitors. WR stars account for approximately $45$ per cent of BH-forming progenitors under this prescription, rather than forming the majority, although hot blue progenitors remain dominant overall, as WR and BSG stars together constitute approximately $60$ per cent of the BH-forming population. Fig.\,\ref{fig:cmd_Ertl} also includes ultra-stripped progenitors classified as non-exploding by the \citet{Ertl2016} criterion. As these models have total masses below $2\,\mathrm{M}_\odot$, however, failure to launch a successful explosion does not imply BH formation. Instead, these systems would undergo quiet neutron-star formation. While the \citet{Ertl2016} criterion predicts a higher overall failed-collapse rate and a significantly larger contribution from RSGs, the broad conclusion that a large fraction of BH-forming progenitors are hot and blue remains unchanged.

\label{lastpage}

\end{document}